\documentclass[%
reprint,
frontmatterverbose,
preprintnumbers,
nobibnotes,
amsmath,amssymb,
aps,
prc,
]{revtex4-2}
\usepackage[pdftex]{graphicx}
\usepackage{dcolumn}
\usepackage{bm}
\usepackage[pdftex]{color}
\usepackage{CJK}
\usepackage[T1]{fontenc}
\usepackage{mathrsfs}
\def\ket#1{\left|{#1}\right\rangle}
\def\braket#1#2{\left\langle{#1}\middle|{#2}\right\rangle}
\def\brakket#1#2#3{\left\langle{#1}\middle|{#2}\middle|{#3}\right\rangle}
\def\ve#1{{\bm{#1}}}

\def\urm#1{\scriptstyle{\text{\textrm{\textmd{\textup{#1}}}}}}
\def\uurm#1{\scriptscriptstyle{\text{\textrm{\textmd{\textup{#1}}}}}}
\def\avr#1{\left\langle{#1}\right\rangle}

\let\temp\epsilon
\let\epsilon\varepsilon
\let\varepsilon\temp
\let\temp\relax
\let\temp\phi
\let\phi\varphi
\let\varphi\temp
\let\temp\relax
\DeclareMathOperator{\laplace}{\Delta}

\DeclareMathOperator{\softplus}{softplus}
\DeclareMathOperator{\ReLU}{ReLU}
\begin{document}
% 
%%%%%%%%%%%%%%%%%%%%%%%%%%%%%%%%%%%%%%%%%%%%%%%%%% 
\begin{CJK*}{UTF8}{}
  \preprint{RIKEN-iTHEMS-Report-23}
  \preprint{KUNS-2954}
  \title{Multi-body wave function of ground and low-lying excited states using unornamented deep neural networks}
  \author{Tomoya Naito (\CJKfamily{min}{内藤智也})}
  \email{
    tnaito@ribf.riken.jp}
  \affiliation{
    RIKEN Interdisciplinary Theoretical and Mathematical Sciences Program (iTHEMS),
    Wako 351-0198, Japan}
  \affiliation{
    Department of Physics, Graduate School of Science, The University of Tokyo,
    Tokyo 113-0033, Japan}
  \author{Hisashi Naito (\CJKfamily{min}{内藤久資})}
  \email{
    naito@math.nagoya-u.ac.jp}
  \affiliation{
    Graduate School of Mathematics, Nagoya University,
    Nagoya 464-8602, Japan}
  \author{Koji Hashimoto (\CJKfamily{min}{橋本幸士})}
  \email{
    koji@scphys.kyoto-u.ac.jp}
  \affiliation{
    Department of Physics, Kyoto University,
    Kyoto 606-8502, Japan}
  \date{\today}
  %%%%%%%%%%%%%%%%%%%%%%%%%%%%%%%%%%%%%%%%%%%%%%%%%% 
  \begin{abstract}
    We propose a method to calculate wave functions and energies
    not only of the ground state
    but also of low-lying excited states
    using a deep neural network and the unsupervised machine learning technique.
    For systems composed of identical particles,
    a simple method to perform symmetrization for bosonic systems and antisymmetrization for fermionic systems is also proposed.
  \end{abstract}
  \maketitle
\end{CJK*}
%%%%%%%%%%%%%%%%%%%%%%%%%%%%%%%%%%%%%%%%%%%%%%%%%%
%
% Introduction
%
\section{Introduction}
\par
Atoms, molecules, and solids are composed of many electrons and ions,
and 
atomic nuclei are composed of many nucleons.
In principle, once the Schr\"{o}dinger equation of these systems is solved, 
most properties can be described.
However, in practice, they are quantum many-fermion systems,
which are difficult to solve directly. 
Hence, it has been one of the important issues to solve the Schr\"{o}dinger equation for the quantum many-fermion efficiently and accurately;
in fact, many numerical methods, for instance, 
the Faddeev calculation~\cite{
  Faddeev1961Sov.Phys.JETP12_1014}, 
several methods for few-body systems~\cite{
  Yakubovskii1967Sov.J.Nucl.Phys.5_937,
  FabredelaRipelle1983Ann.Phys.147_281,
  Carlson1987Phys.Rev.C36_2026,
  Kamimura1988Phys.Rev.A38_621,
  Kamada1992Nucl.Phys.A548_205,
  Varga1995Phys.Rev.C52_2885,
  Viviani1995Few-BodySyst.18_25,
  Navratil1999Phys.Rev.C59_1906,
  Navratil2000Phys.Rev.C61_044001,
  Barnea2000Phys.Rev.C61_054001,
  Kamada2001Phys.Rev.C64_044001},
the quantum Monte Carlo method (QMC)
including the variational Monte Carlo, diffusion Monte Carlo (DMC), and auxiliary-field quantum Monte Carlo methods~\cite{
  Ceperley1978Phys.Rev.B18_3126,
  Ceperley1980Phys.Rev.Lett.45_566,
  Shepherd2012Phys.Rev.B85_081103,
  Shepherd2012J.Chem.Phys.136_244101,
  Booth2013Nature493_365,
  Lonardoni2018Phys.Rev.C97_044318,
  Shen2020J.Chem.Phys.153_204108},
the configuration interaction method~\cite{
  Pople1976Int.J.QuantumChem.10_1,
  Pople1977Int.J.QuantumChem.12_149,
  Pople1999Rev.Mod.Phys.71_1267},
the coupled cluster method~\cite{
  Coester1958Nucl.Phys.7_421,
  Coester1960Nucl.Phys.17_477,
  Cizek1966J.Chem.Phys.45_4256,
  Cizek1971Int.J.QuantumChem.5_359},
the density functional theory (DFT)~\cite{
  Hohenberg1964Phys.Rev.136_B864,
  Kohn1965Phys.Rev.140_A1133,
  Kohn1999Rev.Mod.Phys.71_1253},
the density matrix renormalization group~\cite{
  White1992Phys.Rev.Lett.69_2863,
  White1999J.Chem.Phys.110_4127,
  Schollwoeck2005Rev.Mod.Phys.77_259,
  Baiardi2020J.Chem.Phys.152_040903},
the dynamical mean-field theory~\cite{
  Anisimov1997J.Phys.Condens.Matter9_7359,
  Kotliar2006Rev.Mod.Phys.78_865,
  Knizia2012Phys.Rev.Lett.109_186404},
and
the lattice effective field theory~\cite{
  Brockmann1992Phys.Rev.Lett.68_1830,  
  Chen2004Phys.Rev.Lett.92_257002,
  Lee2009Prog.Part.Nucl.Phys.63_117,
  Drut2013J.Phys.G40_043101}
have been proposed in recent decades.
\par
Among the above, DFT and QMC are classified into methods based on the variational principle.
The variational principle~\cite{
  Schiff1968QuantumMechanics_McGrawHill}
guarantees that
the ground-state energy $ E_{\urm{gs}} $ of a Hamiltonian $ H $ satisfies 
\begin{equation}
  \label{eq:variational_principle}
  E_{\urm{gs}}
  =
  \inf
  \frac{\brakket{\Psi}{H}{\Psi}}{\braket{\Psi}{\Psi}},
\end{equation}
where all the possible functions are considered in the infimum.
The minimizer corresponds to the ground-state wave function.
Indeed, it is scarcely possible to consider all the possible functions throughout minimizing the energy expectation value;
hence, in practice, the calculation accuracy of a method based on the variational principle depends on the ansatz of a trial wave function.
In other words,
the size of the space of trial wave functions, in principle, determines the calculation accuracy.
\par
For instance, a trial wave function of DFT is a Slater determinant,
which is the simplest antisymmetric trial wave function.
Owing to the simpleness of the ansatz, the numerical cost is drastically reduced,
while it is known that interparticle correlation is partially missing~\cite{
  Becke2014J.Chem.Phys.140_18A301}.
\par
In the QMC calculation, a Jastrow-type trial wave function~\cite{
  Jastrow1955Phys.Rev.98_1479}
is often used.
A Jastrow-type wave function $ \ket{\Psi} $ consists of
a single- (or sometimes multi-) Slater determinant $ \ket{\Phi_0} $ and a correlation factor $ F $,
$ \ket{\Psi} = F \ket{\Phi_0} $.
With assuming $ F $ as a symmetric function, $ \ket{\Psi} $ satisfies antisymmetry.
Owing to the introduction of the factor $ F $,
interparticle correlations are described better than a single Slater determinant.
Nevertheless, most QMC calculations optimize mainly $ F $,
while $ \ket{\Phi_0} $ is optimized only around an initial ansatz~\cite{
  Seth2011J.Chem.Phys.134_084105}.
In addition, an ansatz is introduced even for $ F $;
hence, accuracy also depends on the ansatz.
Recently, based on the QMC calculation,
a deep neural network (DNN) has been used for the ansatz of a trial wave function~\cite{
  Nomura2017Phys.Rev.B96_205152,
  Pfau2020Phys.Rev.Research2_033429,
  Hermann2020Nat.Chem.12_891}.
Since deep neural networks span much wider space,
calculation accuracy is much improved,
which is guaranteed by the universal approximation theorem~\cite{
  Cybenko1989Math.ControlSignalsSyst.2_303,
  Hornik1991NeuralNetw.4_251}.
Nevertheless, once the ansatz of a trial wave function is introduced,
the systematic improvement of the calculation is difficult even with a DNN
since the space trial wave functions span is already limited.
\par
Another problem of variational-principle-based methods is calculation of excitation spectra.
Excitation spectra are important quantities of molecules and atomic nuclei,
while the variational principle [Eq.~\eqref{eq:variational_principle}] obtains the ground state only.
Hence, another technique is needed to calculate excited states based on the variational principle.
Indeed, a method to calculate low-lying excited states using the DMC calculation was proposed~\cite{
  Cheon1996Prog.Theor.Phys.96_971}
by considering the orthogonality of wave functions,
while its application has been still limited~\cite{
  Masui1998Prog.Theor.Phys.100_977},
whereas 
calculation of excited states on top of the DFT ground state
has been widely performed by using the random phase approximation or some other techniques~\cite{
  Hedin1999J.Phys.Condens.Matter11_R489,
  Bender2003Rev.Mod.Phys.75_121,
  Colo2011J.Phys.Conf.Ser.321_012018,
  Nakatsukasa2016Rev.Mod.Phys.88_045004,
  Reining2018WIREsComput.Mol.Sci.8_e1344},
while there is still room to be improved.
Recently, low-lying excited states were also obtained in Ref.~\cite{
  Entwistle2023Nat.Commun.14_274}
using the combination of the DNN, QMC, and the orthogonal condition.
\par
In this paper, we propose a new method to calculate energies and wave functions of the ground state and low-lying excited states based on the variational principle.
The ground-state wave function is assumed to be a DNN,
which, in principle, is able to represent any function~\cite{
  Cybenko1989Math.ControlSignalsSyst.2_303,
  Hornik1991NeuralNetw.4_251}.
Using an essence of the machine learning technique---the minimization of the loss function---,
they are directly optimized by using the machine learning technique.
In the machine learning technique, 
a network structure is introduced and the parameters of the network are optimized 
with minimizing the loss function.
This process is often called training.
In many works, these parameters are trained by using the given data called learning data,
whose size is often large.
On the contrary, the machine leaning trained without learning data,
called unsupervised machine learning,
has also been used in many different fields.
References~\cite{
  Saito2018J.Phys.Soc.Japan87_074002,
  Keeble2020Phys.Lett.B809_135743}
also assumed the ground-state wave function as a deep neural network
that were optimized by using the machine learning technique:
Ref.~\cite{
  Saito2018J.Phys.Soc.Japan87_074002}
performed calculation of few-body bosonic systems
and
Ref.~\cite{
  Keeble2020Phys.Lett.B809_135743}
performed calculation of the simplest realistic system---a deuteron.
Although these papers are pioneering works of unsupervised machine learnings for quantum many-body problems,
the fermion antisymmetrization was not considered
and
excited states were not studied,
while both ground and excited states of many-fermionic systems are interesting in general.
Recently, Ref.~\cite{
  2022arXiv220912572W}
proposed a method to obtain the ground state of the many-body Schr\"{o}dinger equation for fermionic systems
by using a tensor neural network,
while its implementation is involved
and
the antisymmetrization is, indeed, not perfectly guaranteed.
\par
In this paper, on top of the method in Refs.~\cite{
  Saito2018J.Phys.Soc.Japan87_074002,
  Keeble2020Phys.Lett.B809_135743}
a simple method of the antisymmetrization for many-fermion systems
or the symmetrization for many-boson systems
is introduced.
Then, low-lying excited states are sequentially calculated by using the orthogonality conditions and the variational principles.
In this method,
there is no need to discover a DNN architecture to generate (anti)symmetric wave functions.
The (anti)symmetrization is put at the level of loss functions.
Furthermore, the symmetrization and the antisymmetrization are implemented in almost the same way
and
the wave function perfectly satisfies (anti)symmetry.
Thanks to the simplicity of the implementation, the numerical cost is quite small.
We show that our method works successfully for popular examples in bosonic and fermionic quantum mechanical systems,
providing a fundamental basis of the DNN method for quantum mechanics.
\par
This paper is organized as follows:
Section~\ref{sec:ground_state} is devoted to calculation of the ground-state.
The novel (anti)symmetrization is introduced.
Section~\ref{sec:excited_state} is devoted to the calculation of low-lying excited states.
All the calculations are performed in a MacBook Pro with the Apple M1 chip (MacBook Pro (13-inch, M1, 2020): MacBookPro17,1) and $ 16 \, \mathrm{GB} $ memory.
Section~\ref{sec:summary} gives a summary of this paper.
% 
% Ground State
%
\section{Ground-State Calculation}
\label{sec:ground_state}
\par
In this section, the ground-state wave function and energy are calculated by using a DNN and the machine learning technique.
Throughout the paper, a machine learning software named \textsc{Tensorflow}~\cite{
  tensorflow2015-whitepaper}
is used.
\subsection{Network structure and machine learning technique}
\label{subsec:network}
\par
In general, a wave function of a $ d $-dimensional $ N $-particle system is a function of the spatial coordinates of all the particles
$ \ve{r}_j = \left( r_{j1}, r_{j2}, \ldots, r_{jd} \right) $
($ j = 1 $, $ 2 $, \ldots, $ N $).
Here, for the sake of simplicity, we neglect the spin and isospin dependence of wave functions
and
$ \ve{R} $ denotes
$ \ve{R}
= \left( \ve{r}_1, \ve{r}_2, \ldots, \ve{r}_N \right)
= \left( r_{11}, r_{12}, \ldots, r_{1d}, r_{21}, r_{22}, \ldots, r_{2d}, \ldots, r_{N1}, r_{N2}, \ldots, r_{Nd} \right) $.
\par
\begin{figure}[tb]
  \centering
  \includegraphics[width=1.0\linewidth]{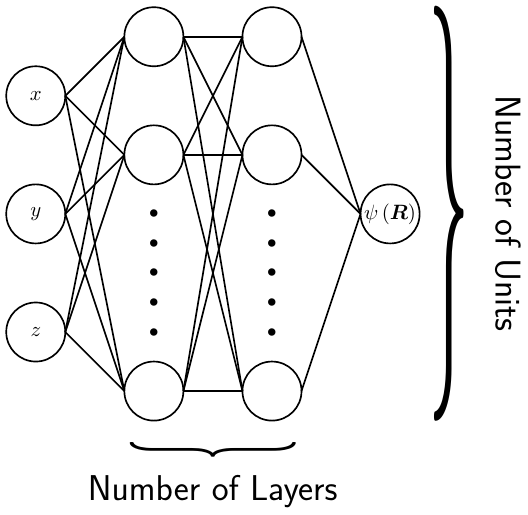}
  \caption{
    Schematic figure of the deep neural network representing a one-dimensional three-body system.}
  \label{fig:schematic_dnn}
\end{figure}

In this work, the wave function is represented by a deep neural network
with $ Nd $-input units that corresponds to the spatial coordinate $ \ve{R} $
and
one-output unit that corresponds to the value of the wave function $ \psi \left( \ve{R} \right) $.
Between the input and output layers, there are hidden layers.
Each unit is connected to all the units just one before or after layers.
The schematic figure of the deep neural network is shown in Fig.~\ref{fig:schematic_dnn}.
In this paper, the ``softplus'' function
\begin{equation}
  \softplus \left( x \right)
  =
  \log
  \left( 1 + e^x \right)
\end{equation}
is used for an activation function
and
the Adam optimizer~\cite{
  2014arXiv1412.6980K}
is used for the optimization process.
\par
As the normal procedure of the numerical calculation of the wave function,
the spatial coordinate is discretized.
Each point is treated as a batch of the machine learning.
In other words, if the spatial coordinate of each direction is discretized with $ M $ meshes,
the batch size is the same as the number of meshes, $ M^{dN} $.
The mini-batch technique is not used.
\par
Once the spatial coordinates are discretized, 
the Hamiltonian 
\begin{equation}
  \label{eq:Hamiltonian}
  H
  =
  -
  \frac{\hbar^2}{2m}
  \sum_j 
  \laplace_j
  +
  \sum_j
  V^{\urm{ext}} \left( \ve{r}_j \right)
  +
  \frac{1}{2}
  \sum_{j \ne k}
  V^{\urm{int}} \left( \ve{r}_j, \ve{r}_k \right)
\end{equation}
can be written as a matrix,
where
$ m $ is the mass of the particles,
$ V^{\urm{ext}} $ is the external potential,
and
$ V^{\urm{int}} $ is the interparticle interaction.
The matrices of the external potential and the interaction are diagonal
and that of the kinetic energy is sparse.
Hence, the expectation value of the Hamiltonian $ \avr{H} $ can be calculated by using the sparse-matrix technique.
The ground-state wave function minimizes $ \avr{H} $;
therefore, $ \avr{H} $ is regarded as a loss function.
Note that all the calculations are performed with double precision floating point numbers (\texttt{float64}).
For simplicity, $ m = \hbar = 1 $ is assumed.
\par
The procedure in the \textsc{Tensorflow} code is as follows:
\begin{enumerate}
\item Construct a model of the deep neural network;
\item Note that although a \textsc{Tensorflow} subroutine specification for the loss function technically requires two inputs---the training data (\texttt{true\_value})
  and the network output (\texttt{predicts}),
  the former is not referred in our training;
\item Fit the model (\texttt{model.fit})
  where the initial value of \texttt{predicts} consists of positive random numbers;  
\item The final wave function \texttt{output\_wf} is given by using \texttt{model.predict};
\item The ground-state energy is calculated using the wave function obtained by the last step.
\end{enumerate}
The third step (\texttt{model.fit}) corresponds to determining the parameters inside the DNN;
the fourth step (\texttt{model.predict}) corresponds to storing the wave function obtained in the previous step;
the fifth step corresponds to calculating the ground-state energy using the wave function obtained in the fourth step.
Note that \texttt{predicts} and the final wave function should be normalized whenever generated.
\subsection{One-dimensional one-particle systems}
\label{subsec:1d_1body}
\par
In this section, benchmark calculations of one-dimensional systems are shown.
The dependence of the numbers of units and layers on calculation accuracy is also discussed.
Since there exists only one particle,
there is no interaction, $ V^{\urm{int}} \equiv 0 $;
thus, the Hamiltonian reads
\begin{equation}
  H
  =
  -
  \frac{1}{2}
  \frac{d^2}{dx^2}
  +
  V^{\urm{ext}} \left( x \right).
\end{equation}
\par
Since we focus only on bound states in this paper,
it is enough to deal with the limited spatial region.
In the calculation,
$ \left| x \right| \le x_{\urm{max}} $ is considered and the box is discretized within $ 1024 $ meshes, i.e., $ M = 1024 $.
The Dirichlet boundary condition
($ \psi \left( \pm x_{\urm{max}} \right) = 0 $) is used.
For the second derivative, the three-point derivative is used for simplicity,
while it can be straightforwardly improved for the accuracy~\cite{
  Tajima2001Prog.Theor.Phys.Suppl.142_265}.
Then, $ H $ is discretized as
\begin{subequations}    
  \begin{align}
    H
    \simeq
    \tilde{H}
    & =
      -
      \frac{1}{2 h^2}
      \tilde{T}
      +
      \tilde{V}^{\urm{ext}}, \\
    \tilde{T}
    & = 
      \begin{pmatrix}
        -2 & 1  & 0  & \ldots & 0 & 0 & 0 \\
        1  & -2 & 1  & \ldots & 0 & 0 & 0 \\
        0  & 1  & -2 & \ldots & 0 & 0 & 0 \\
        \vdots & \vdots & \vdots & \ddots & \vdots & \vdots & \vdots \\
        0 & 0 & 0 & \ldots & -2 &  1 & 0  \\
        0 & 0 & 0 & \ldots &  1 & -2 & 1  \\
        0 & 0 & 0 & \ldots &  0 &  1 & -2 \\
      \end{pmatrix}, \\
    \tilde{V}^{\urm{ext}}
    & = 
      \begin{pmatrix}
        V^{\urm{ext}}_1 & 0 & 0 & \ldots & 0 & 0 & 0 \\
        0 & V^{\urm{ext}}_2 & 0 & \ldots & 0 & 0 & 0 \\
        0 & 0 & V^{\urm{ext}}_3 & \ldots & 0 & 0 & 0 \\
        \vdots & \vdots & \vdots & \ddots & \vdots & \vdots & \vdots \\
        0 & 0 & 0 & \ldots & V^{\urm{ext}}_{M - 3} & 0 & 0 \\
        0 & 0 & 0 & \ldots & 0 & V^{\urm{ext}}_{M - 2} & 0 \\
        0 & 0 & 0 & \ldots & 0 & 0 & V^{\urm{ext}}_{M - 1} \\
      \end{pmatrix}
  \end{align}
\end{subequations}
and the wave function is also discretized as a $ \left( M - 1 \right) $-dimensional vector
\begin{equation}
  \psi
  \simeq
  \tilde{\psi}
  =
  \begin{pmatrix}
    \psi_1 \\
    \psi_2 \\
    \psi_3 \\
    \vdots \\
    \psi_{M - 3} \\
    \psi_{M - 2} \\
    \psi_{M - 1} 
  \end{pmatrix},
\end{equation}
where $ \tilde{\psi} $ is assumed to be normalized,
i.e.,
$ h \sqrt{\sum_j \tilde{\psi}_j^2} = 1 $,
$ V^{\urm{ext}}_j = V^{\urm{ext}} \left( x_j \right) $,
$ \psi_j = \psi \left( x_j \right) $,
$ x_j = - x_{\urm{max}} + hj $,
and 
$ h $ denotes the mesh size $ h = 2 x_{\urm{max}} / M $~\footnote{
  If $ \tilde{H} $ is directly discretized,
  wave functions of the ground state and $ \left( M - 2 \right) $ excited states can be obtained.}.
This $ \tilde{\psi} $ is used for \texttt{predicts} and \texttt{output\_wf}.
Here, a tilde denotes discretized form.
Then,
$ \avr{H} $ can be calculated as
\begin{equation}
  \avr{H}
  \simeq
  \tilde{\avr{H}}
  =
  \tilde{\psi}^{\mathsf{T}} \tilde{H} \tilde{\psi}.
\end{equation}
\subsubsection{Harmonic oscillator}
\label{subsubsec:1d_1body_ho}
\par
First of all, the harmonic oscillator potential
\begin{equation}
  \label{eq:pot_harmonic}
  V^{\urm{ext}} \left( x \right)
  =
  \frac{1}{2}
  \omega^2 x^2 
\end{equation}
is tested.
The ground-state wave function $ \psi_{\urm{gs}} $ and energy $ E_{\urm{gs}} $ are,
respectively, known exactly as~\cite{
  Schiff1968QuantumMechanics_McGrawHill}
\begin{subequations}
  \begin{align}
    \psi_{\urm{gs}} \left( x \right)
    & =
      \left(
      \frac{\omega}{\pi}
      \right)^{1/4}
      \exp
      \left(
      - \frac{\omega x^2}{2}
      \right),
      \label{eq:1d_1body_ho_exact_wf} \\
    E_{\urm{gs}}
    & =
      \frac{\omega}{2}.
      \label{eq:1d_1body_ho_exact_energy}
  \end{align}
\end{subequations}
In this calculation, $ x_{\urm{max}} = 5 $ is used.
\par
Table~\ref{tab:ho} shows the summary of calculations.
In general, all the calculations give almost the correct energy [Eq.~\eqref{eq:1d_1body_ho_exact_energy}].
On the one hand, total optimization costs similar amount of time in all the calculation.
On the other hand, different setup requires different number of epochs and time per epoch for optimizing the DNN.
Small DNN tends to take shorter time for each epoch,
while it requires longer epochs.
It seems that $ 32 $ units per layer is too large,
so it requires longer epoch and longer estimation time per epoch.
It should be noted that the number of epochs differs in each run
since the initial condition of the fitting procedure is generated by the random numbers.
In addition, if one uses a different value of the learning rate,
the number of epochs can be different.
\par
\begin{figure*}[tb]
  \centering
  \begin{minipage}{0.32\linewidth}
    \centering
    \includegraphics[width=1.0\linewidth]{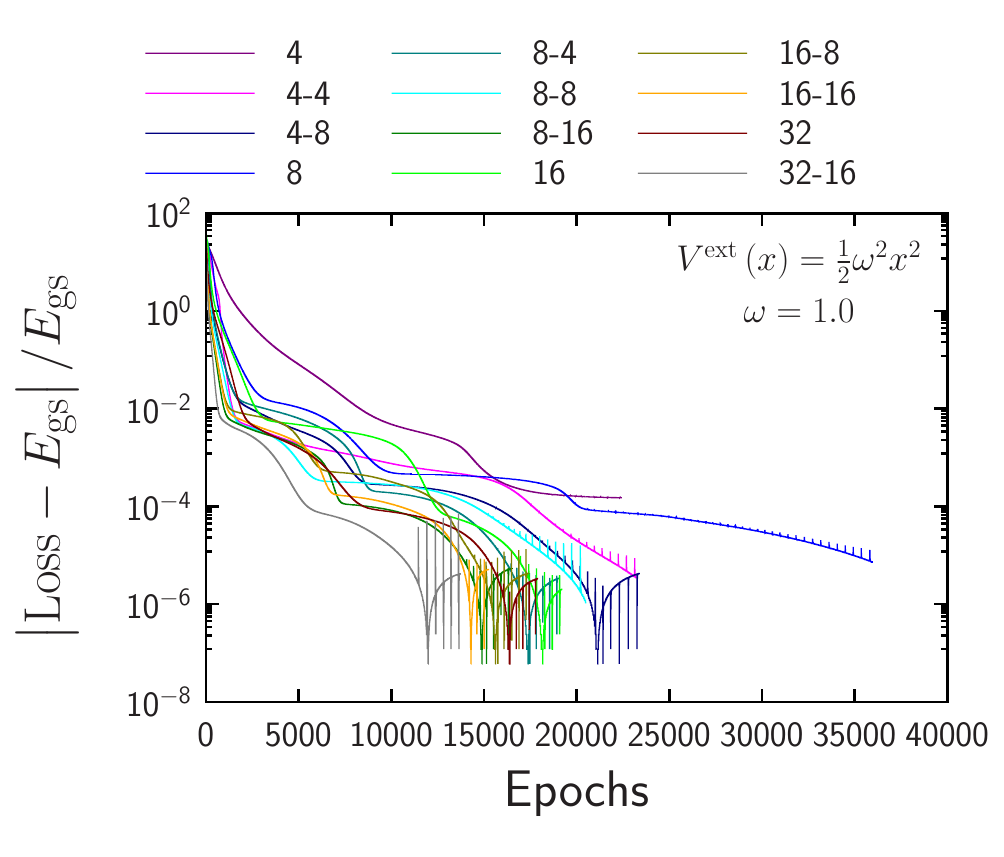}
  \end{minipage}
  \hfill
  \begin{minipage}{0.32\linewidth}
    \centering
    \includegraphics[width=1.0\linewidth]{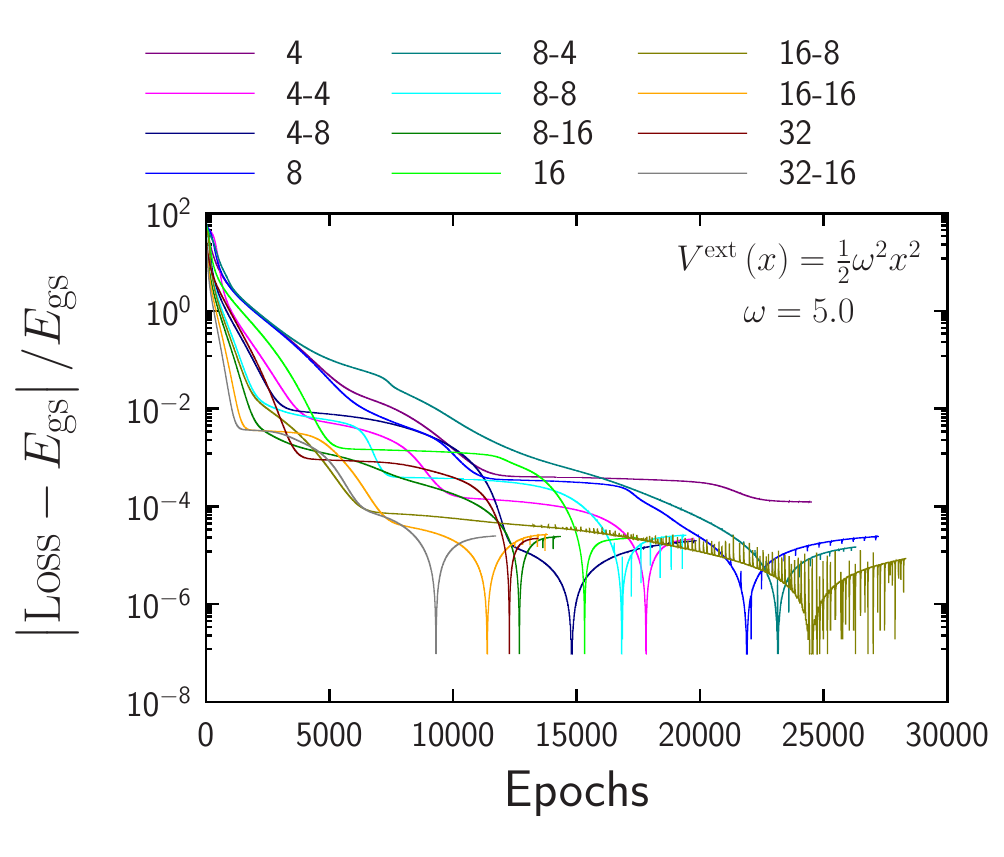}
  \end{minipage}
  \hfill
  \begin{minipage}{0.32\linewidth}
    \centering
    \includegraphics[width=1.0\linewidth]{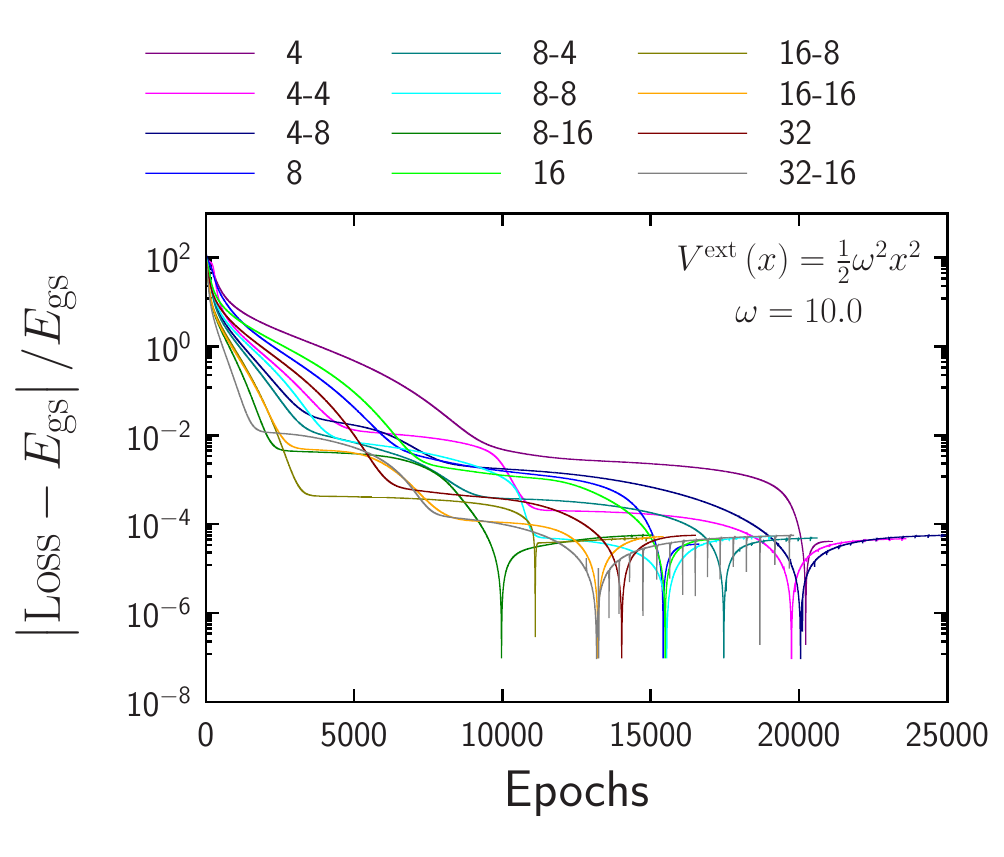}
  \end{minipage}
  \caption{
    Relative error of $ \avr{H} $ to the exact ground-state energy $ E_{\urm{gs}} $
    for the harmonic oscillator potential
    as functions of the number of epochs.}
  \label{fig:ho_loss}
\end{figure*}

Figure~\ref{fig:ho_loss} shows relative errors of the loss function, $ \avr{H} $, to the exact ground-state energy $ E_{\urm{gs}} $ as functions of the number of epochs.
It can be seen that, although the loss function achieved the relative error of $ 1.0 \times 10^{-8} $,
the final accuracy becomes about $ 1.0 \times 10^{-4} $.
This may be due to the precision of the \textsc{Tensorflow} code.
\par
\begin{figure*}[tb]
  \centering
  \begin{minipage}{0.32\linewidth}
    \centering
    \includegraphics[width=1.0\linewidth]{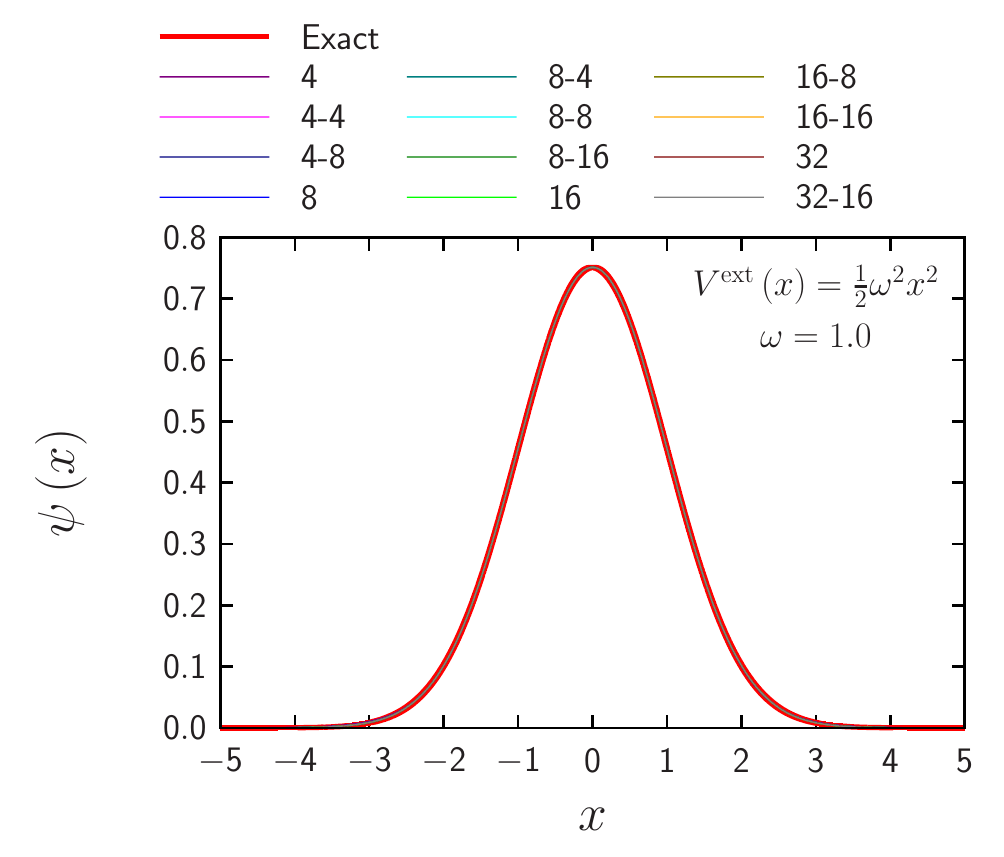}
  \end{minipage}
  \hfill
  \begin{minipage}{0.32\linewidth}
    \centering
    \includegraphics[width=1.0\linewidth]{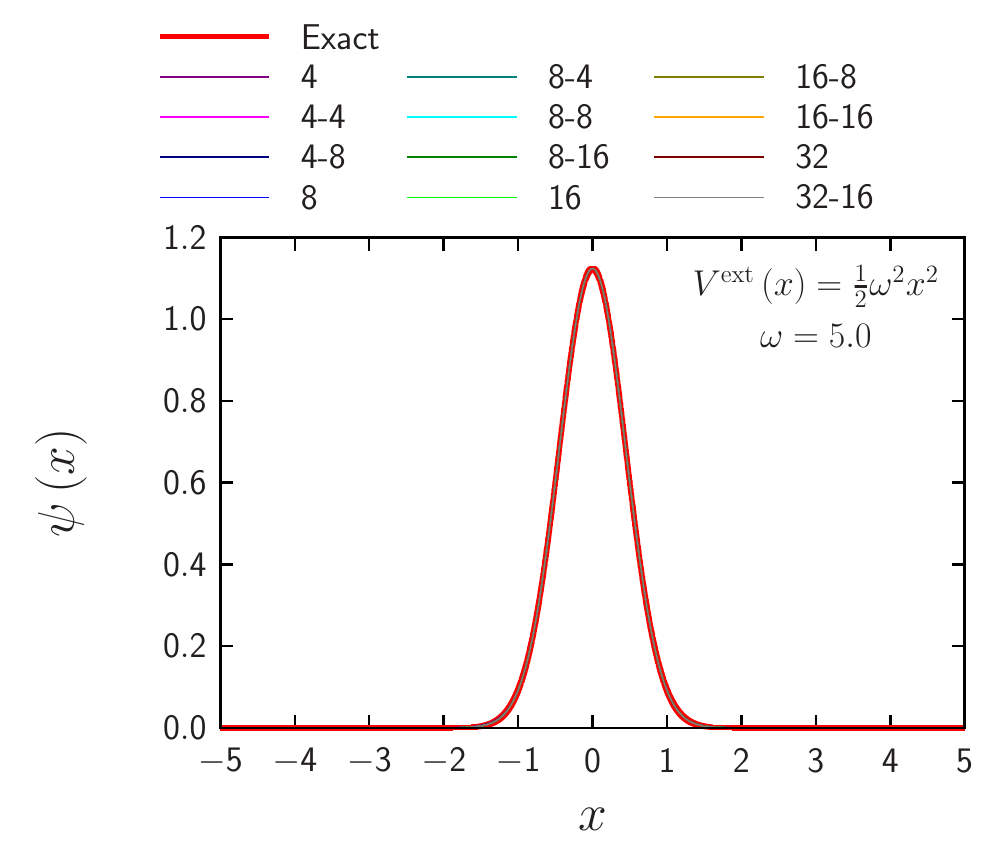}
  \end{minipage}
  \hfill
  \begin{minipage}{0.32\linewidth}
    \centering
    \includegraphics[width=1.0\linewidth]{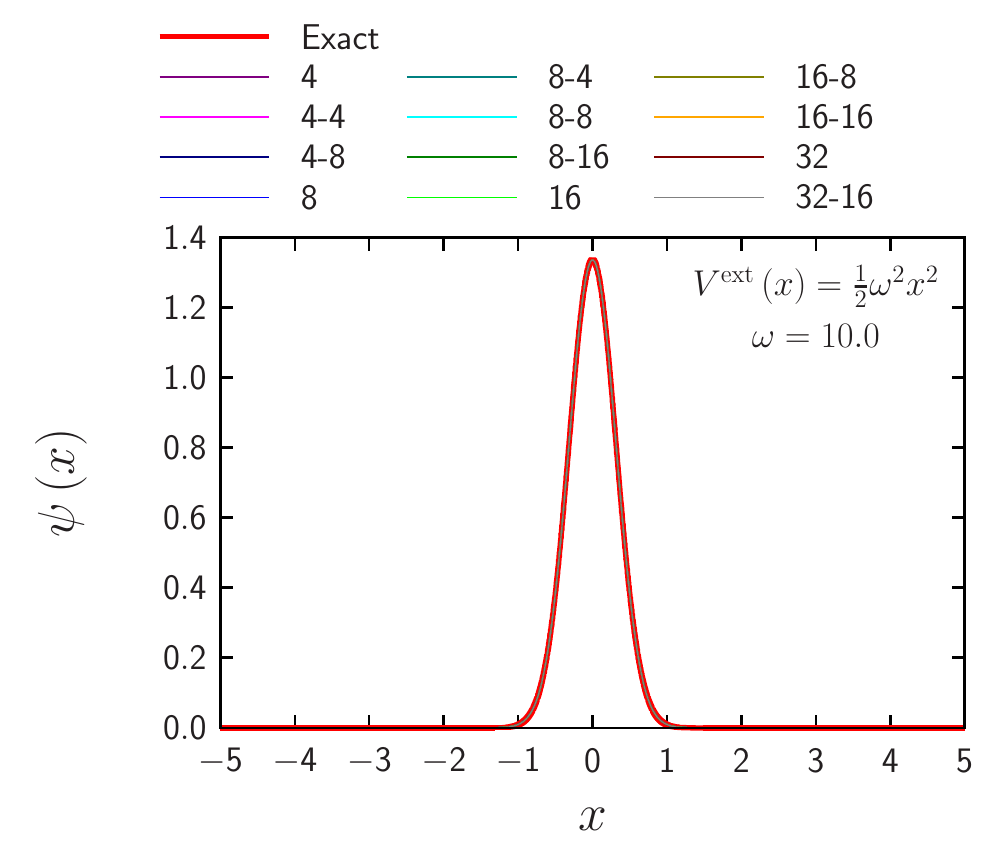}
  \end{minipage}
  \caption{
    Wave function under the harmonic oscillator potential.
    The red thick line corresponds to the exact solution [Eq.~\eqref{eq:1d_1body_ho_exact_wf}],
    while thin lines correspond to results of DNN calculation.
    Different thin line corresponds to different number of units.
    We observe that all the simulated results overlap with the exact solution.}
  \label{fig:ho_wf}
\end{figure*}
\begin{figure*}[tb]
  \centering
  \begin{minipage}{0.32\linewidth}
    \centering
    \includegraphics[width=1.0\linewidth]{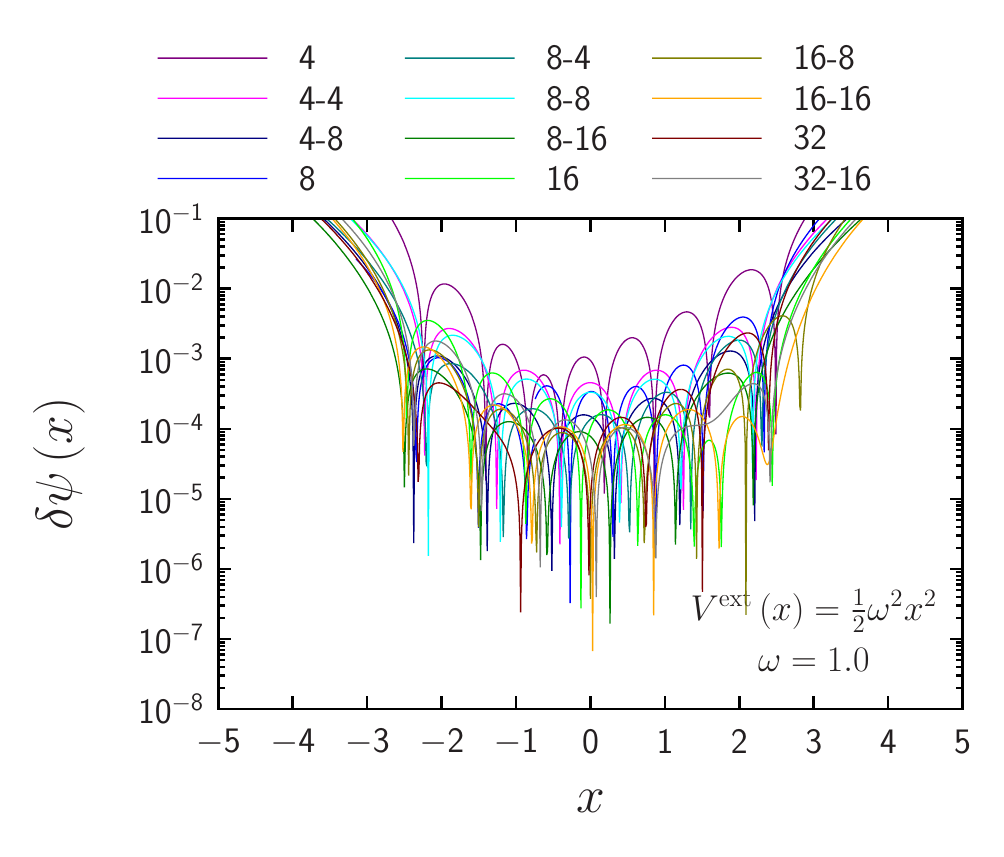}
  \end{minipage}
  \hfill
  \begin{minipage}{0.32\linewidth}
    \centering
    \includegraphics[width=1.0\linewidth]{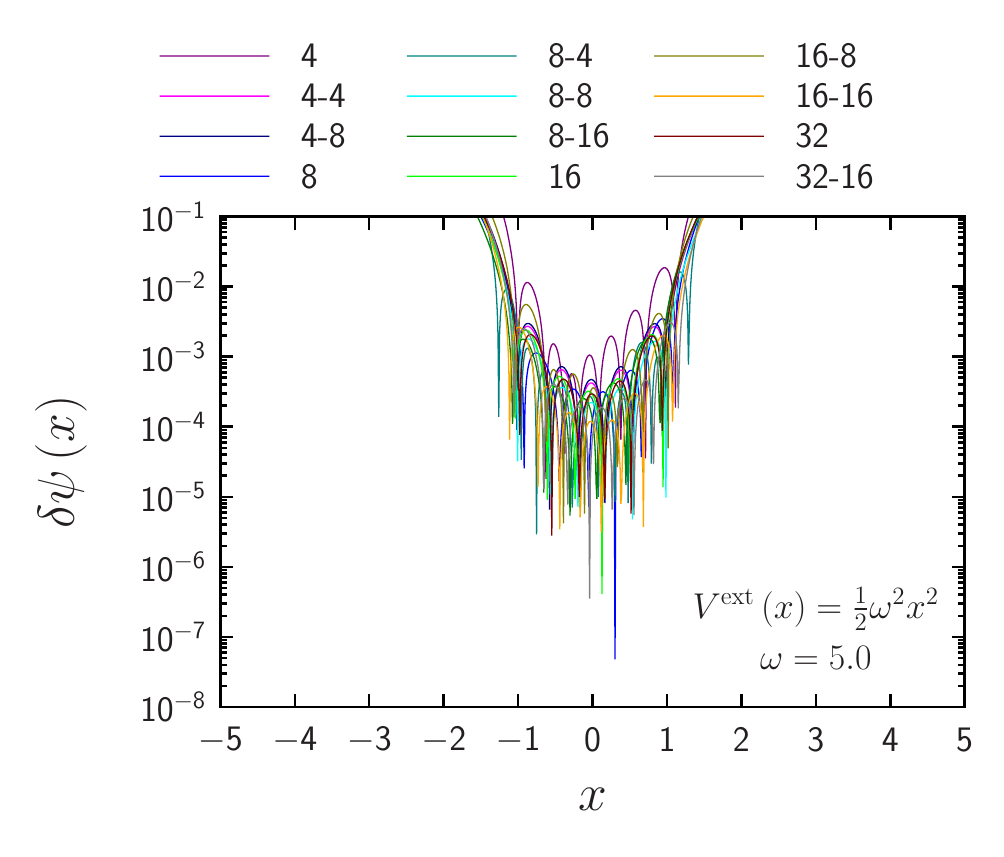}
  \end{minipage}
  \hfill
  \begin{minipage}{0.32\linewidth}
    \centering
    \includegraphics[width=1.0\linewidth]{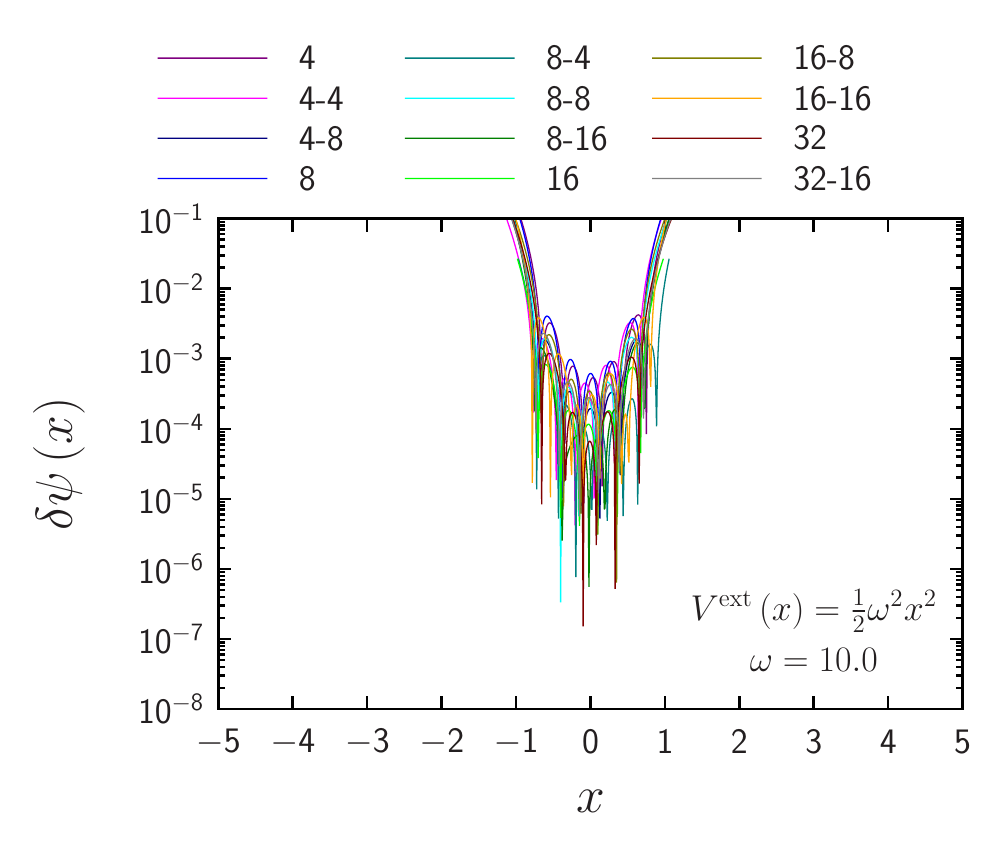}
  \end{minipage}
  \caption{
    Relative error of DNN wave function to exact one for the harmonic oscillator potential.
    Different thin line corresponds to different number of units.
    In the region of large $ \left| x \right| $,
    the deviation $ \delta \psi \left( x \right) $ diverges,
    because the denominator of Eq.~\eqref{eq:wf_deviation},
    $ \psi^{\urm{exact}} \left( x \right) $, reaches to zero.
    Hence, the figures plot only $ \left| \delta \phi \left( x \right) \right| < 10^{-1} $.}
  \label{fig:ho_wf_error}
\end{figure*}

Figure~\ref{fig:ho_wf} shows calculated wave functions.
The red thick lines correspond to the exact solution given in Eq.~\eqref{eq:1d_1body_ho_exact_wf},
while thin lines correspond to the results given in this work,
where different colors correspond to different numbers of units and layers.
The relative errors of the DNN wave function, $ \psi^{\urm{DNN}} $, to the exact one, $ \psi^{\urm{exact}} $,
\begin{equation}
  \label{eq:wf_deviation}
  \delta \psi \left( x \right)
  =
  \frac{\left| \psi^{\urm{DNN}} \left( x \right) - \psi^{\urm{exact}} \left( x \right) \right|}{\psi^{\urm{exact}} \left( x \right)} 
\end{equation}
are shown in Fig.~\ref{fig:ho_wf_error}.
It can be seen that the DNN calculation, basically, reproduces the exact solution
in our interest 
within the accuracy of $ 10^{-4} $ or more.
This deviation can be reduced if we use more tight convergence criterion~\footnote{
  Due to the internal code of the \textsc{Tensorflow},
  when the error of the loss function is calculated,
  a single precision floating point number (\texttt{float32}) seems to be used.
  Thus, the error of the loss function becomes zero if its actual value is smaller than about $ 1.0 \times 10^{-7} $.
  Note that the convergence criteria of the DFT calculation is often about $ 1.0 \times 10^{-8} $ or even $ 1.0 \times 10^{-10} $,
  for instance, see the sample input of Ref.~\cite{
    ADPACK}.}.
In the tail region, the deviation $ \delta \psi \left( x \right) $ diverges,
while this is because the denominator of Eq.~\eqref{eq:wf_deviation},
$ \psi^{\urm{exact}} \left( x \right) $, reaches to zero.
The deviation looks larger if $ \omega $ is smaller,
which is related to the cutoff parameter for the spatial mesh $ x_{\urm{min}} $.
The exact value of $ \psi_{\urm{gs}} \left( x_{\urm{min}} \right) $
is $ 2.8 \times 10^{-6} $ for $ \omega = 1.0 $,
while it is $ 8.1 \times 10^{-28} $ for $ \omega = 5.0 $
and it is much smaller for $ \omega = 10.0 $,
while in the numerical calculation, they are approximated to zero.
The value $ 2.8 \times 10^{-6} $ may be too large to assume to be zero.
\par
It should be noted that rather small DNN is enough to reproduce the solution of the wave function.
Owing to the simplicity, it is easy to analyze the weights and biases of the DNN.
For instance, the DNN wave function for the single layer with four units
includes only $ 13 $ parameters;
the ground-state DNN wave function for $ \omega = 1.0 $ can be written as
\begin{widetext}
  \begin{subequations}
    \begin{align}
      \psi_{\urm{gs}} \left( x \right)
      & =
        \frac{1}{3.7451}
        \softplus
        \left(
        a_{\urm{gs}} \left( x \right)
        \right),
        \label{eq:wf_ho_dnn} \\
      a_{\urm{gs}} \left( x \right)
      & = 
        2.4069 a_1 \left( x \right)
        -
        1.8344 a_2 \left( x \right)
        -
        1.9778 a_3 \left( x \right)
        +
        2.3484 a_4 \left( x \right)
        -
        4.8998, 
        \label{eq:wf_ho_dnn_a} \\
      a_1 \left( x \right)
      & =
        \softplus
        \left( 
        0.35953 x
        +
        3.9226
        \right), \\
      a_2 \left( x \right)
      & =
        \softplus
        \left(
        2.5821 x
        +
        0.033213
        \right), \\
      a_3 \left( x \right)
      & =
        \softplus
        \left(
        -0.65170 x 
        +
        2.9574
        \right), \\
      a_4 \left( x \right)
      & =
        \softplus
        \left(
        0.15421 x 
        +
        2.2016
        \right),
    \end{align}
  \end{subequations}
\end{widetext}
where
the first coefficient of Eq.~\eqref{eq:wf_ho_dnn} ($ 1/3.7451 $) is not obtained by the DNN
but instead by the normalization.
Hence, smaller DNN is better not only due to the calculation cost but also for analysis of the structure of DNN.
\par
\begin{figure}[tb]
  \centering
  \includegraphics[width=1.0\linewidth]{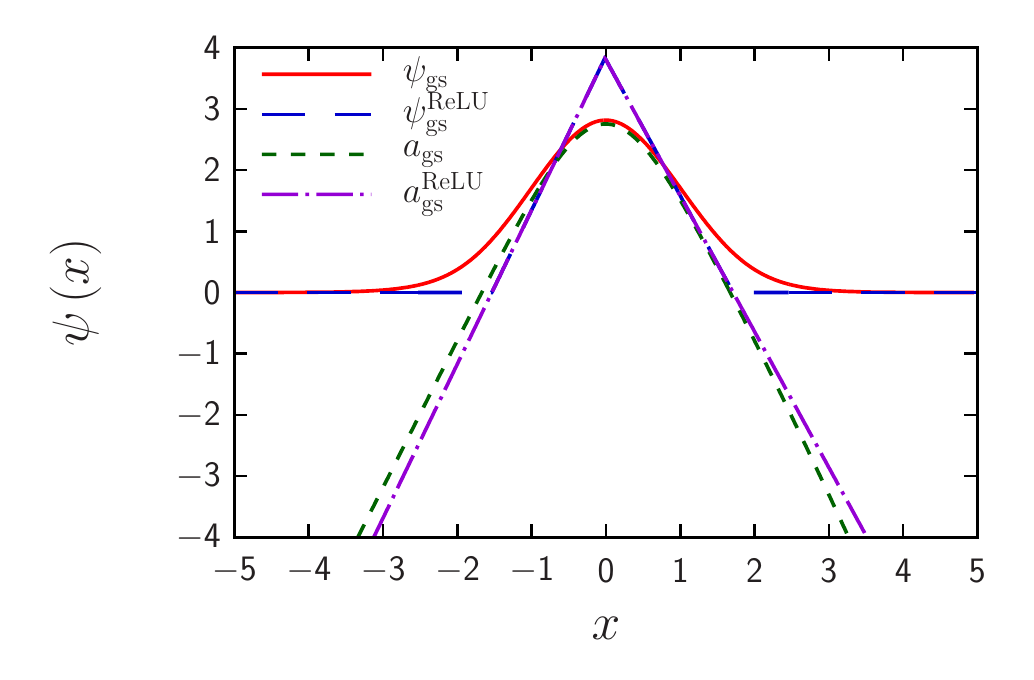}
  \caption{
    Wave function obtained by the DNN [Eq.~\eqref{eq:wf_ho_dnn}]
    and the obtained function before the output layer [Eq.~\eqref{eq:wf_ho_dnn_a}],
    where the normalization is ignored.
    Equations \eqref{eq:wf_ho_dnn} and \eqref{eq:wf_ho_dnn_a}
    where the softplus function is replaced with the ReLU function
    are also plotted as $ \psi_{\urm{gs}}^{\urm{ReLU}} $ and $ a_{\urm{gs}}^{\urm{ReLU}} $, respectively.}
  \label{fig:dnn_analysis_1d_ho}
\end{figure}

Let us provide our interpretation of the obtained wave function [Eq.~\eqref{eq:wf_ho_dnn}].
The rectified linear function (ReLU)
\begin{equation}
  \label{eq:relu}
  \ReLU \left( x \right)
  =
  \begin{cases}
    0 & \text{($ x < 0 $)}, \\
    x & \text{($ x \ge 0 $)}
  \end{cases}
\end{equation}
is a widely used activation function,
and the softplus function can be regarded as a smoothed version of the ReLU.
Here, for interpreting Eq.~\eqref{eq:wf_ho_dnn},
we shall just replace the softplus function with the ReLU.
The wave function obtained by the DNN [Eq.~\eqref{eq:wf_ho_dnn}]
and the obtained function before the output layer [Eq.~\eqref{eq:wf_ho_dnn_a}]
are shown in Fig.~\ref{fig:dnn_analysis_1d_ho},
where the normalization factor ($ 1/3.7451 $) of $ \psi_{\urm{gs}} $ is ignored.
Equations \eqref{eq:wf_ho_dnn} and \eqref{eq:wf_ho_dnn_a}
where the softplus function is replaced to the ReLU function
are also plotted as $ \psi_{\urm{gs}}^{\urm{ReLU}} $ and $ a_{\urm{gs}}^{\urm{ReLU}} $, respectively.
The DNN with the ReLU function can be understood as an approximation with a piecewise linear function.
As shown in Fig.~\ref{fig:dnn_analysis_1d_ho},
the ground-state wave function in DNN is approximated by the following function:
\begin{equation}
  \psi_{\urm{gs}}
  \approx
  \begin{cases}
    -ax + b & \text{($ 0 \le x \le b/a $)}, \\
    a x + b & \text{($ -b/a \le x \le 0 $)}, \\
    0       & \text{(otherwise)},
  \end{cases}
\end{equation}
where $ a $ and $ b $ are positive numbers.
The ReLU function at the output layer guarantees
to make the wave function vanish for $ x < -b/a $ and $ x > b/a $,
and thus,
$ a_{\urm{gs}} $ should be $ \pm ax + b $.
This function can be represented by just two ReLU functions.
Hence, even two units in hidden layer are enough to describe the brief structure of ground-state wave function,
and
with increasing the number of units, the ground-state wave function is reproduced easily.
Since the ReLU function is not differentiable at $ x = 0 $,
the ReLU wave function is not differentiable.
Hence, the softplus is better to describe a differentiable function,
while the ReLU function can also describe a differentiable function approximately if the number of units is large enough.
In case of $ N $-bodies systems, 
the similar function to Eq.~\eqref{eq:wf_ho_dnn} can be represented by just $ 2^N $ ReLU functions.
\begin{table*}[tb]
  \centering
  \caption{
    Calculation summary of a one-body problem under the harmonic oscillator potential.
    Row with ``---'' in the column ``\# of unit for 2nd layer'' corresponds to
    calculation performed only with one layer.}
  \label{tab:ho}
  \begin{ruledtabular}
    \begin{tabular}{dddddddd}
      \multicolumn{1}{c}{$ \omega $} & \multicolumn{2}{c}{\# of Unit} & \multicolumn{3}{c}{Energy} & \multicolumn{1}{c}{\# of Epochs} & \multicolumn{1}{c}{Time per Epoch} \\
      \cline{2-3}
      \cline{4-6}
                                     & \multicolumn{1}{c}{1st Layer} & \multicolumn{1}{c}{2nd Layer} & \multicolumn{1}{c}{Kinetic} & \multicolumn{1}{c}{Potential} & \multicolumn{1}{c}{Total} & & \multicolumn{1}{c}{($ \mathrm{\mu s} $)} \\
      \hline
      1.0 & 4 & \multicolumn{1}{c}{---} & +0.250043 & +0.250032 & +0.500075 & 22448 & 474.909 \\
      1.0 & 4 & 4 & +0.250006 & +0.249996 & +0.500002 & 23242 & 542.892 \\
      1.0 & 4 & 8 & +0.250001 & +0.249997 & +0.499998 & 23408 & 588.101 \\
      1.0 & 8 & \multicolumn{1}{c}{---} & +0.250002 & +0.250002 & +0.500004 & 35973 & 511.120 \\
      1.0 & 8 & 4 & +0.250000 & +0.249998 & +0.499998 & 19072 & 592.471 \\
      1.0 & 8 & 8 & +0.250004 & +0.249996 & +0.500001 & 20531 & 657.621 \\
      1.0 & 8 & 16 & +0.249999 & +0.249999 & +0.499997 & 16540 & 725.032 \\
      1.0 & 16 & \multicolumn{1}{c}{---} & +0.250000 & +0.249999 & +0.499999 & 19239 & 566.930 \\
      1.0 & 16 & 8 & +0.250000 & +0.249998 & +0.499998 & 17425 & 724.952 \\
      1.0 & 16 & 16 & +0.249999 & +0.249998 & +0.499997 & 15275 & 869.706 \\
      1.0 & 32 & \multicolumn{1}{c}{---} & +0.250000 & +0.249999 & +0.499998 & 17924 & 687.738 \\   
      1.0 & 32 & 16 & +0.249999 & +0.249999 & +0.499998 & 13761 & 999.868 \\
      \hline
      5.0 & 4 &  \multicolumn{1}{c}{---} &+1.250192 & +1.250116 & +2.500308 & 24517 & 458.700 \\
      5.0 & 4 & 4 & +1.250025 & +1.249921 & +2.499946 & 19772 & 551.123 \\
      5.0 & 4 & 8 & +1.250027 & +1.249923 & +2.499951 & 19855 & 605.380 \\
      5.0 & 8 & \multicolumn{1}{c}{---} & +1.249994 & +1.249946 & +2.499939 & 27232 & 514.425 \\
      5.0 & 8 & 4 & +1.250453 & +1.249510 & +2.499963 & 26324 & 630.199 \\
      5.0 & 8 & 8 & +1.249994 & +1.249942 & +2.499936 & 19440 & 661.411 \\
      5.0 & 8 & 16 & +1.250002 & +1.249937 & +2.499939 & 14364 & 731.111 \\
      5.0 & 16 & \multicolumn{1}{c}{---} & +1.249967 & +1.249975 & +2.499942 & 17423 & 568.478 \\
      5.0 & 16 & 8 & +1.250041 & +1.249938 & +2.499979 & 28345 & 731.765 \\
      5.0 & 16 & 16 & +1.249977 & +1.249957 & +2.499934 & 13851 & 855.756 \\
      5.0 & 32 & \multicolumn{1}{c}{---} & +1.249919 & +1.250026 & +2.499945 & 13461 & 671.493 \\
      5.0 & 32 & 16 & +1.249972 & +1.249966 & +2.499938 & 11745 & 1005 \\
      \hline
      10.0 & 4 & \multicolumn{1}{c}{---} & +2.499927 & +2.499866 & +4.999793 & 21134 & 482.504 \\
      10.0 & 4 & 4 & +2.500071 & +2.499694 & +4.999765 & 23623 & 554.103 \\
      10.0 & 4 & 8 & +2.500000 & +2.499720 & +4.999719 & 24949 & 605.551 \\
      10.0 & 8 & \multicolumn{1}{c}{---} & +2.499877 & +2.499945 & +4.999822 & 16636 & 513.759 \\
      10.0 & 8 & 4 & +2.500038 & +2.499718 & +4.999756 & 20628 & 598.401 \\
      10.0 & 8 & 8 & +2.500003 & +2.499730 & +4.999733 & 19261 & 664.986 \\
      10.0 & 8 & 16 & +2.499947 & +2.499770 & +4.999717 & 15053 & 731.275 \\
      10.0 & 16 & \multicolumn{1}{c}{---} & +2.499873 & +2.499896 & +4.999769 & 17351 & 561.501 \\
      10.0 & 16 & 8 & +2.499996 & +2.499759 & +4.999756 & 14979 & 730.518 \\
      10.0 & 16 & 16 & +2.499917 & +2.499822 & +4.999739 & 15437 & 859.019 \\
      10.0 & 32 & \multicolumn{1}{c}{---} & +2.499834 & +2.499885 & +4.999719 & 16520 & 671.640 \\
      10.0 & 32 & 16 & +2.499976 & +2.499745 & +4.999721 & 19833 & 996.732 \\
    \end{tabular}
  \end{ruledtabular}
\end{table*}
\subsubsection{Square-well potential}
\label{subsubsec:1d_1body_sq}
\par
Next, the square-well potential 
\begin{equation}
  \label{eq:pot_square}
  V^{\urm{ext}} \left( x \right)
  =
  \begin{cases}
    - V_0 & \text{($ \left| x \right| < x_0 $)}, \\
    0     & \text{(otherwise)} 
  \end{cases}
\end{equation}
is tested ($ V_0 > 0 $).
The analytical forms of the ground-state wave function $ \psi_{\urm{gs}} $ and energy $ E_{\urm{gs}} $ are unknown;
thus, our values of the energy will be compared with the numerical calculation obtained by the orthodox method of Hamiltonian diagonalization.
In this calculation, $ x_{\urm{max}} = 20 $ and $ x_0 = 1 $ is used.
\par
Table~\ref{tab:sq} shows the summary of calculations.
In general, all the calculations give almost the correct energy.
The calculation time per epoch and the number of epochs
with respect to the number of layers and units is slightly longer than the case of the harmonic oscillator.
\par
\begin{figure*}[tb]
  \centering
  \begin{minipage}{0.49\linewidth}
    \centering
    \includegraphics[width=1.0\linewidth]{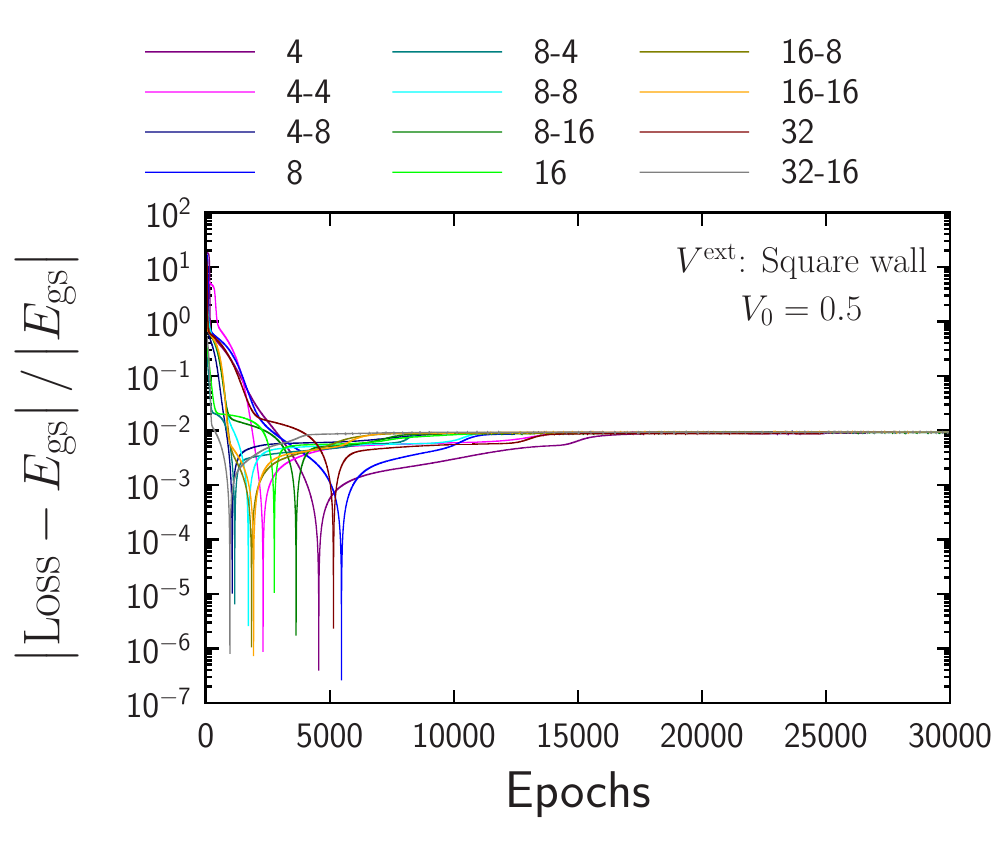}
  \end{minipage}
  \hfill
  \begin{minipage}{0.49\linewidth}
    \centering
    \includegraphics[width=1.0\linewidth]{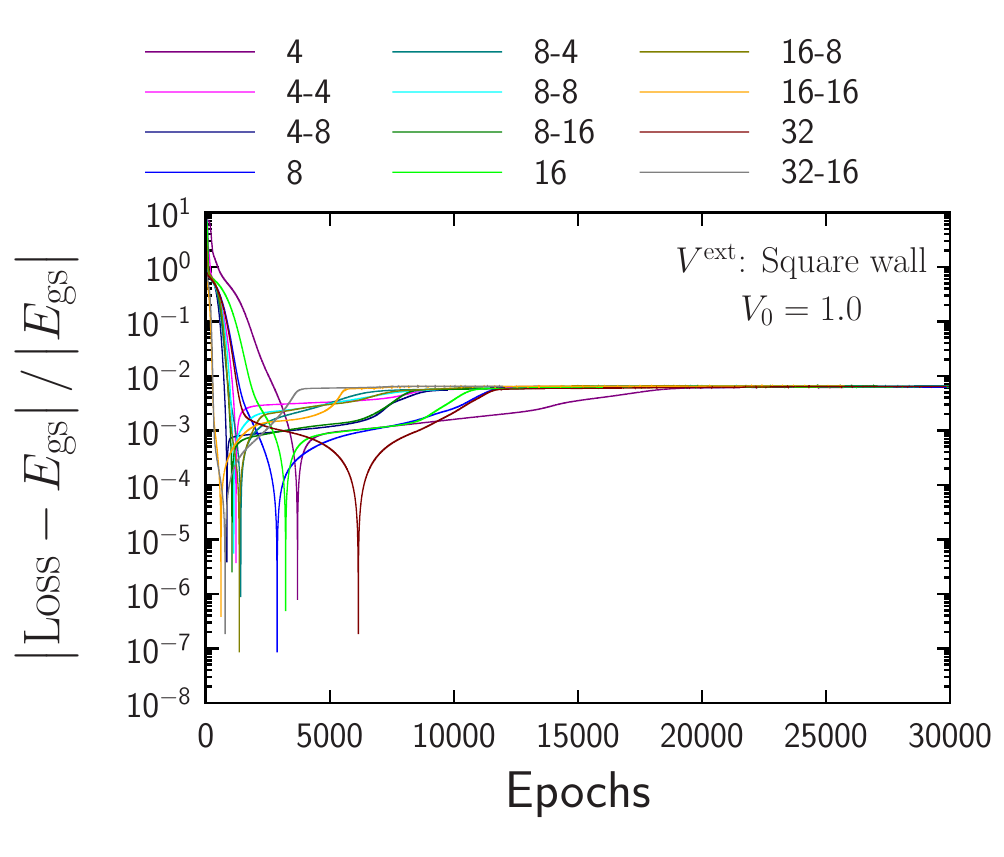}
  \end{minipage}
  \begin{minipage}{0.49\linewidth}
    \centering
    \includegraphics[width=1.0\linewidth]{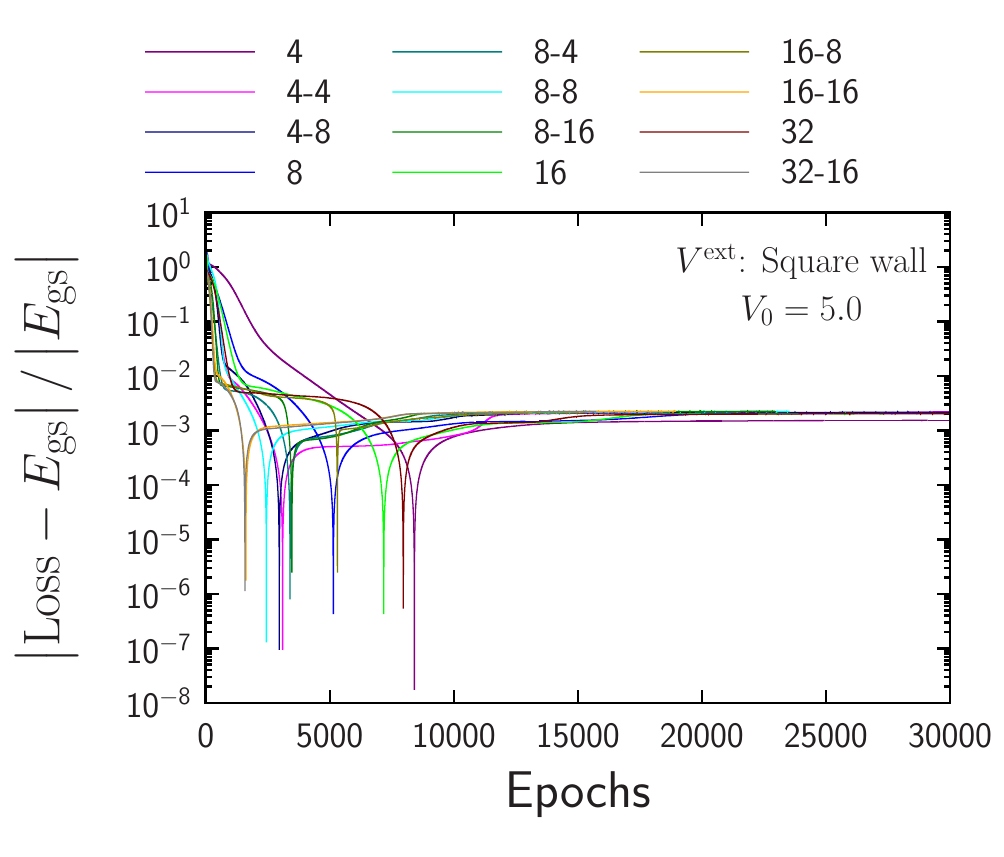}
  \end{minipage}
  \hfill
  \begin{minipage}{0.49\linewidth}
    \centering
    \includegraphics[width=1.0\linewidth]{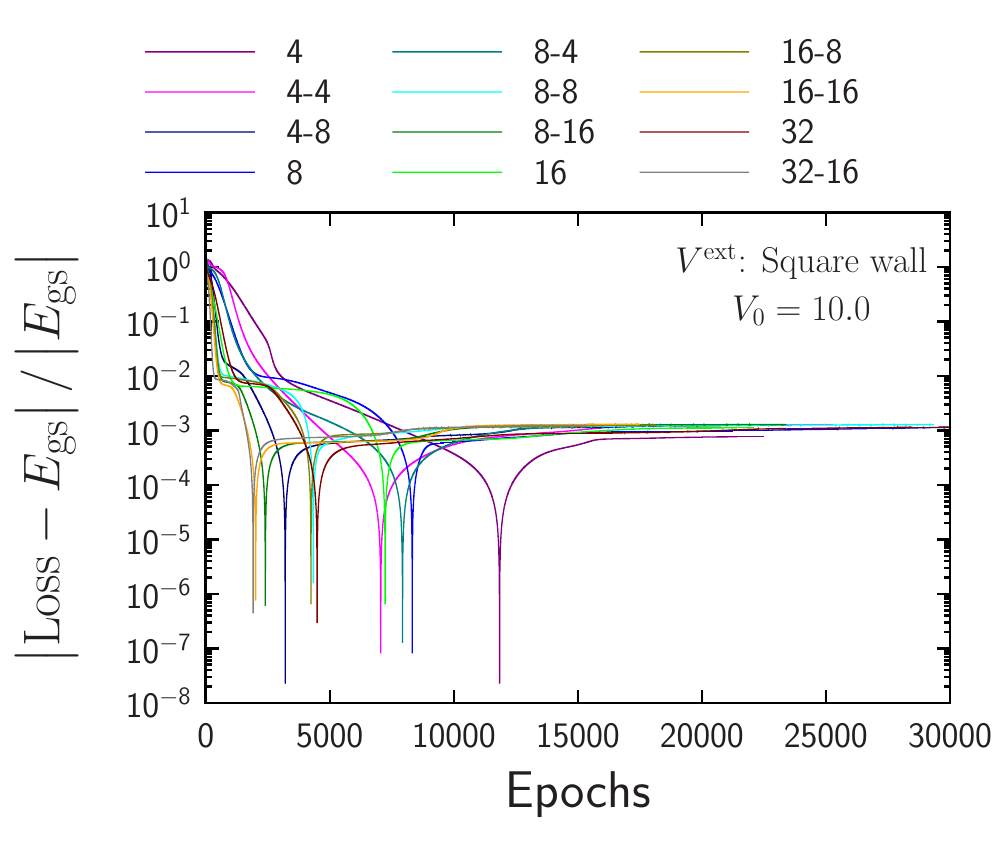}
  \end{minipage}
  \caption{
    Relative error of $ \avr{H} $ to the exact ground-state energy $ E_{\urm{gs}} $
    for the square-well potential
    as functions of the number of epochs.}
  \label{fig:sq_loss}
\end{figure*}

Figure~\ref{fig:sq_loss} shows relative errors of the loss function, $ \avr{H} $, to the exact ground-state energy $ E_{\urm{gs}} $ as functions of the number of epochs.
It can be seen that, although the loss function achieved the relative error of $ 1.0 \times 10^{-7} $,
the final accuracy becomes about $ 1.0 \times 10^{-2} $.
Note that in this calculation, the convergence criteria is needed to be set looser than the other case;
otherwise, it could not reach convergence.
This may be related to the shape of the potential:
Asymptotic region of the square-well potential is zero,
while the harmonic oscillator potential increases rapidly.
It will be shown later that the double-well potential, which is close to the latter situation,
reaches convergence with the tight criterion.
\par
\begin{figure*}[tb]
  \centering
  \begin{minipage}{0.49\linewidth}
    \centering
    \includegraphics[width=1.0\linewidth]{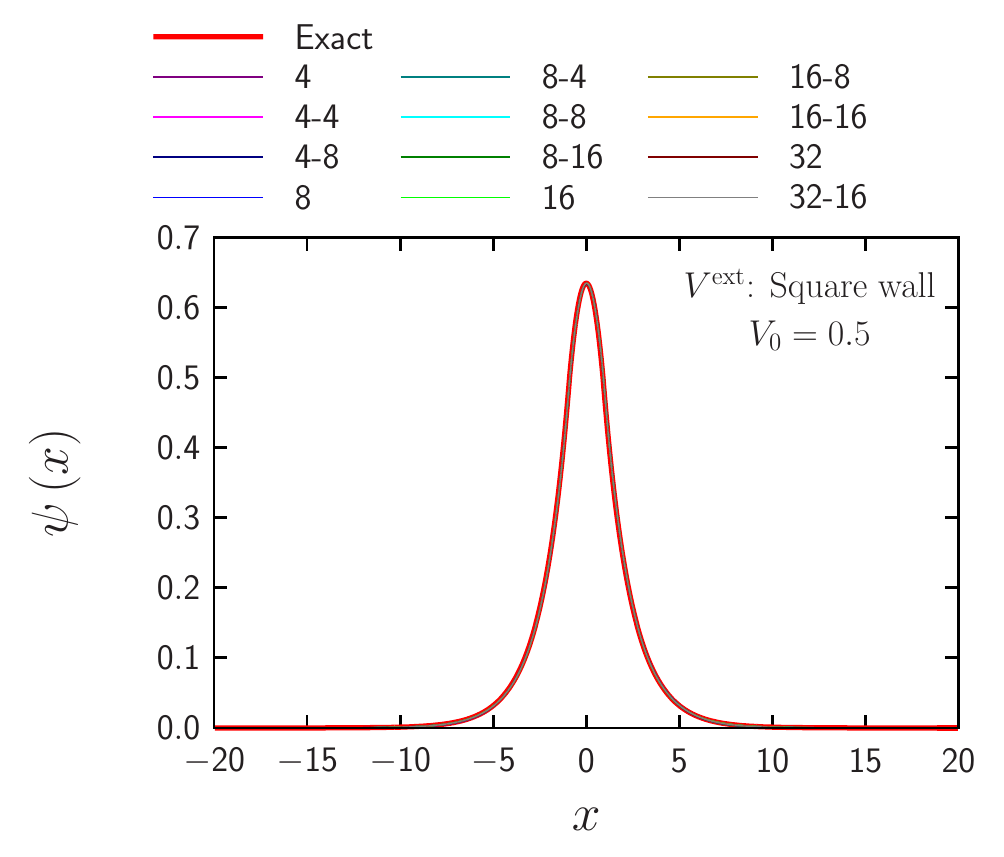}
  \end{minipage}
  \hfill
  \begin{minipage}{0.49\linewidth}
    \centering
    \includegraphics[width=1.0\linewidth]{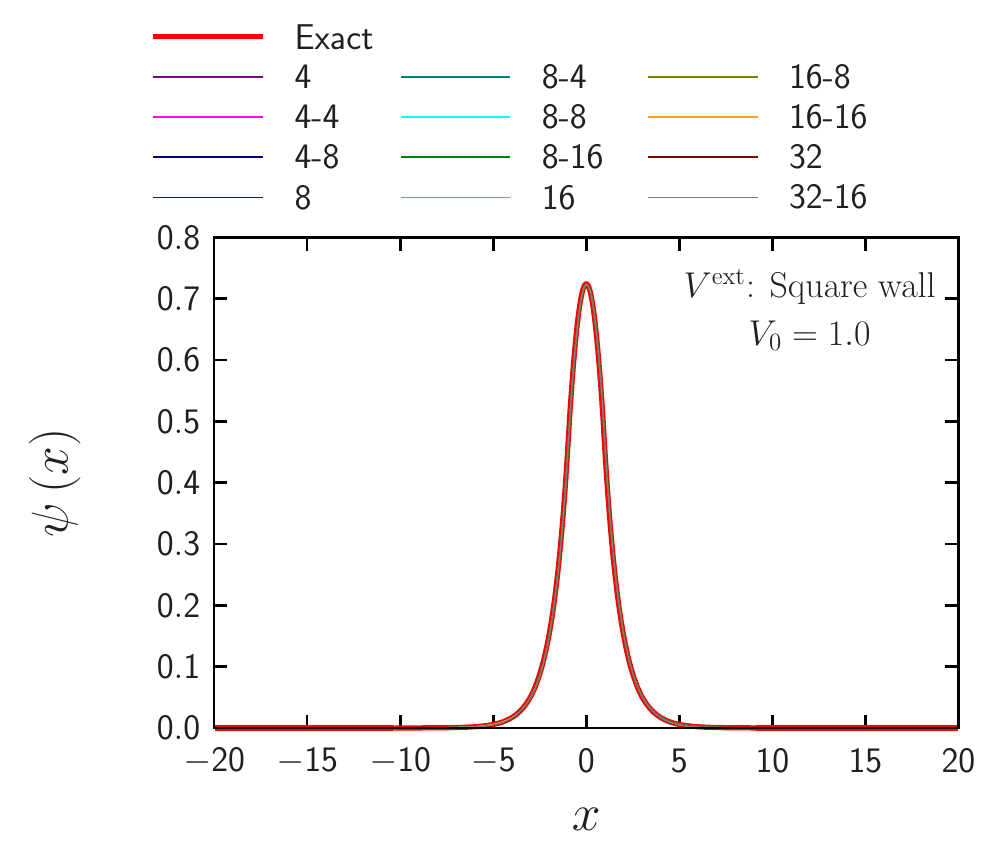}
  \end{minipage}
  \begin{minipage}{0.49\linewidth}
    \centering
    \includegraphics[width=1.0\linewidth]{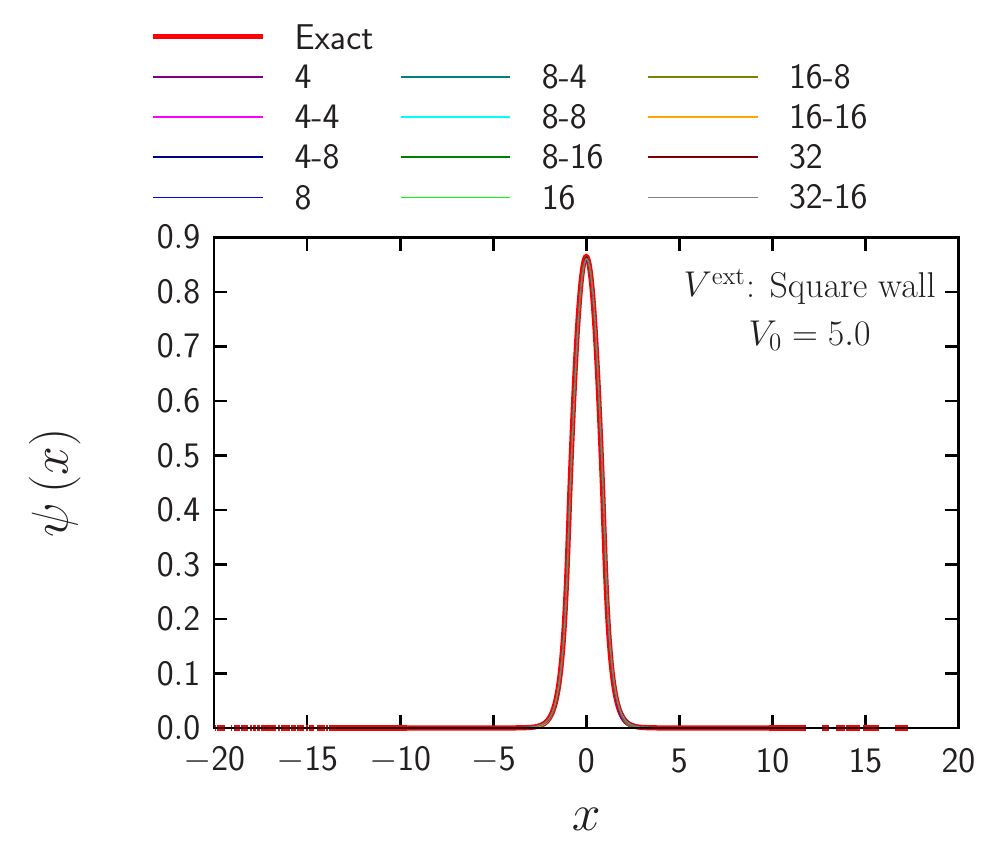}
  \end{minipage}
  \hfill
  \begin{minipage}{0.49\linewidth}
    \centering
    \includegraphics[width=1.0\linewidth]{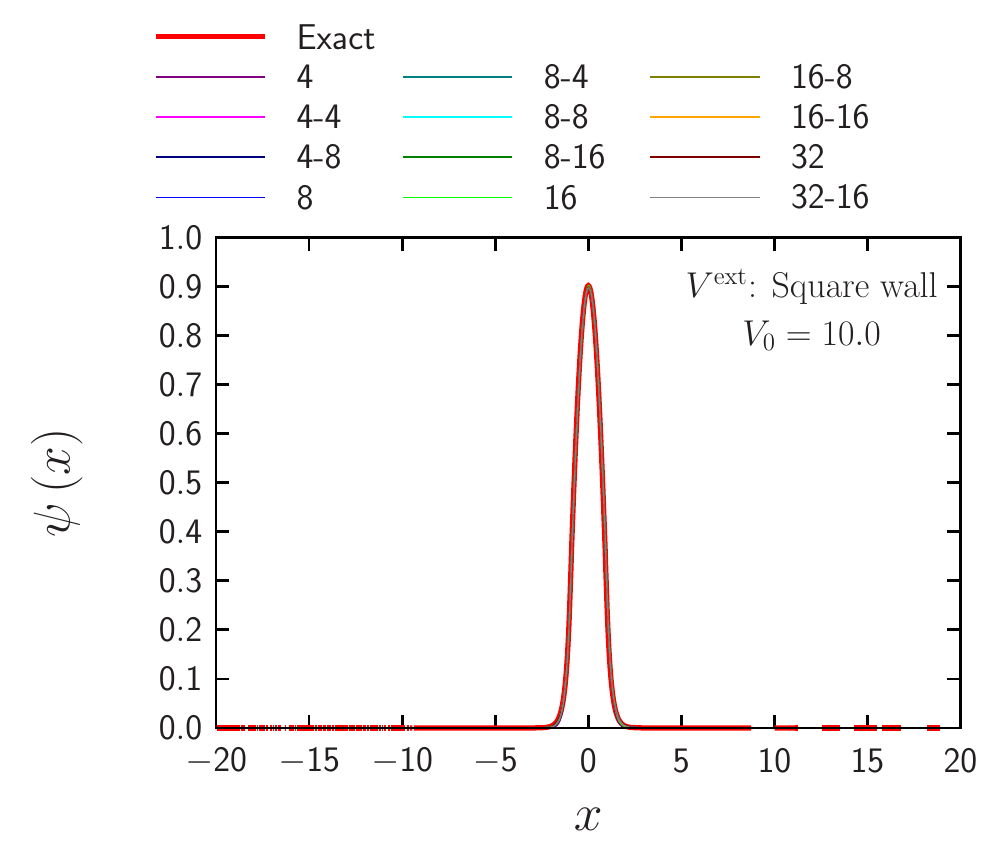}
  \end{minipage}
  \caption{
    Wave function under the square-well potential.
    The red thick line corresponds to the exact solution obtained by the exact diagonalization,
    while thin lines correspond to results of DNN calculation.
    Different thin line corresponds to different number of units.
    Almost all lines overlap with each other.}
  \label{fig:sq_wf}
\end{figure*}

Figure~\ref{fig:sq_wf} shows calculated wave functions.
The red thick lines correspond to the exact solution given by the exact diagonalization,
where the same mesh size matrix form are used, for comparison;
thin lines correspond to the results given in this work.
It can be seen that the DNN calculation, basically, reproduces the solutions given by the exact diagonalization.
\begin{table*}[tb]
  \centering
  \caption{
    Calculation summary of an one-body problem under the square-well potential.
    Row with ``---'' in the column ``\# of unit for 2nd layer'' corresponds to
    calculation performed only with one layer.}
  \label{tab:sq}
  \begin{ruledtabular}
    \begin{tabular}{dddddddd}
      \multicolumn{1}{c}{$ V_0 $} & \multicolumn{2}{c}{\# of Unit} & \multicolumn{3}{c}{Energy} & \multicolumn{1}{c}{\# of Epochs} & \multicolumn{1}{c}{Time per Epoch} \\
      \cline{2-3}
      \cline{4-6}
                                  & \multicolumn{1}{c}{1st Layer} & \multicolumn{1}{c}{2nd Layer} & \multicolumn{1}{c}{Kinetic} & \multicolumn{1}{c}{Potential} & \multicolumn{1}{c}{Total} & & \multicolumn{1}{c}{($ \mathrm{\mu s} $)} \\
      \hline
      0.5 & 4 & \multicolumn{1}{c}{---} & +0.109852 & -0.337864 & -0.228012 & 26600 & 579.787 \\
      0.5 & 4 & 4 & +0.109935 & -0.338093 & -0.228159 & 44065 & 717.730 \\
      0.5 & 4 & 8 & +0.109944 & -0.338110 & -0.228166 & 27771 & 778.485 \\
      0.5 & 8 & \multicolumn{1}{c}{---} & +0.109852 & -0.337876 & -0.228024 & 21802 & 619.806 \\
      0.5 & 8 & 4 & +0.109962 & -0.338088 & -0.228126 & 37964 & 776.147 \\
      0.5 & 8 & 8 & +0.110009 & -0.338173 & -0.228164 & 36410 & 856.319 \\
      0.5 & 8 & 16 & +0.109968 & -0.338119 & -0.228151 & 69731 & 968.769 \\
      0.5 & 16 & \multicolumn{1}{c}{---} & +0.109923 & -0.338026 & -0.228104 & 17015 & 734.647 \\
      0.5 & 16 & 8 & +0.109967 & -0.338116 & -0.228148 & 22157 & 1005 \\
      0.5 & 16 & 16 & +0.109983 & -0.338146 & -0.228163 & 32462 & 1155 \\
      0.5 & 32 & \multicolumn{1}{c}{---} & +0.109847 & -0.337872 & -0.228025 & 24879 & 921.208 \\
      0.5 & 32 & 16 & +0.109976 & -0.338136 & -0.228160 & 34549 & 1400 \\
      \hline
      1.0 & 4 & \multicolumn{1}{c}{---} & +0.206715 & -0.812787 & -0.606072 & 35062 & 572.641 \\
      1.0 & 4 & 4 & +0.206882 & -0.813194 & -0.606312 & 24588 & 720.511 \\
      1.0 & 4 & 8 & +0.206896 & -0.813222 & -0.606326 & 21877 & 787.156 \\
      1.0 & 8 & \multicolumn{1}{c}{---} & +0.206844 & -0.813106 & -0.606261 & 45145 & 609.218 \\
      1.0 & 8 & 4 & +0.206877 & -0.813182 & -0.606305 & 42888 & 769.973 \\
      1.0 & 8 & 8 & +0.207083 & -0.813407 & -0.606325 & 26445 & 834.225 \\
      1.0 & 8 & 16 & +0.206900 & -0.813220 & -0.606320 & 27073 & 976.154 \\
      1.0 & 16 & \multicolumn{1}{c}{---} & +0.206848 & -0.813109 & -0.606261 & 41642 & 735.100 \\
      1.0 & 16 & 8 & +0.206907 & -0.813235 & -0.606328 & 19472 & 1002 \\
      1.0 & 16 & 16 & +0.206888 & -0.813202 & -0.606314 & 25746 & 1155 \\
      1.0 & 32 & \multicolumn{1}{c}{---} & +0.206877 & -0.813112 & -0.606235 & 34259 & 916.159 \\
      1.0 & 32 & 16 & +0.206872 & -0.813146 & -0.606273 & 11987 & 1410 \\
      \hline
      5.0 & 4 & \multicolumn{1}{c}{---} & +0.521370 & -4.821182 & -4.299812 & 45997 & 565.210 \\
      5.0 & 4 & 4 & +0.521033 & -4.823422 & -4.302390 & 34568 & 722.257 \\
      5.0 & 4 & 8 & +0.520966 & -4.823344 & -4.302378 & 38875 & 774.016 \\
      5.0 & 8 & \multicolumn{1}{c}{---} & +0.520887 & -4.822751 & -4.301865 & 32216 & 610.063 \\
      5.0 & 8 & 4 & +0.520913 & -4.823243 & -4.302329 & 17236 & 766.967 \\
      5.0 & 8 & 8 & +0.520972 & -4.823484 & -4.302512 & 23526 & 829.876 \\
      5.0 & 8 & 16 & +0.520893 & -4.823451 & -4.302558 & 22945 & 971.763 \\
      5.0 & 16 & \multicolumn{1}{c}{---} & +0.520781 & -4.822461 & -4.301680 & 27332 & 740.434 \\
      5.0 & 16 & 8 & +0.521167 & -4.823316 & -4.302150 & 15343 & 995.692 \\
      5.0 & 16 & 16 & +0.520927 & -4.823470 & -4.302542 & 18961 & 1151 \\
      5.0 & 32 & \multicolumn{1}{c}{---} & +0.520812 & -4.822494 & -4.301681 & 30846 & 917.315 \\
      5.0 & 32 & 16 & +0.520893 & -4.823362 & -4.302470 & 15731 & 1413 \\
      \hline
      10.0 & 4 & \multicolumn{1}{c}{---} & +0.663220 & -9.846544 & -9.183324 & 22506 & 568.630 \\
      10.0 & 4 & 4 & +0.659302 & -9.847228 & -9.187926 & 25288 & 719.068 \\
      10.0 & 4 & 8 & +0.659456 & -9.846967 & -9.187511 & 20766 & 785.202 \\
      10.0 & 8 & \multicolumn{1}{c}{---} & +0.659267 & -9.846237 & -9.186970 & 34537 & 610.037 \\
      10.0 & 8 & 4 & +0.659308 & -9.847178 & -9.187871 & 23178 & 762.927 \\
      10.0 & 8 & 8 & +0.659146 & -9.847090 & -9.187944 & 29346 & 822.300 \\
      10.0 & 8 & 16 & +0.659161 & -9.847202 & -9.188041 & 23407 & 978.041 \\
      10.0 & 16 & \multicolumn{1}{c}{---} & +0.659727 & -9.846250 & -9.186523 & 22091 & 735.663 \\
      10.0 & 16 & 8 & +0.659073 & -9.846888 & -9.187815 & 18146 & 955.435 \\
      10.0 & 16 & 16 & +0.659103 & -9.847266 & -9.188162 & 17500 & 1153 \\
      10.0 & 32 & \multicolumn{1}{c}{---} & +0.659058 & -9.846543 & -9.187485 & 54691 & 917.252 \\
      10.0 & 32 & 16 & +0.659331 & -9.846907 & -9.187576 & 15394 & 1396 \\
    \end{tabular}
  \end{ruledtabular}
\end{table*}
\subsection{One-dimensional many-particle systems}
\label{subsec:1d_2body}
\par
When one considers systems composed of many identical particles,
the symmetrization for bosonic systems or the antisymmetrization for fermionic systems of the wave function must be considered.
The ground state of the bosonic system is identical to that of the different particles;
hence it has no extra difficulty as was done in Ref.~\cite{
  Saito2018J.Phys.Soc.Japan87_074002},
while the antisymmetrization is rather difficult.
In this section, a simple method of (anti)symmetrization in the DNN wave function is provided,
in which the symmetrization and the antisymmetrization can be performed with equal footing.
\subsubsection{Hamiltonian matrix}
\label{subsubsec:1d_2body_hamil}
\par
As was done in the last section,
the discretized Hamiltonian $ \tilde{H} $ should be represented in a matrix form
and
the discretized wave function $ \tilde{\psi} $ should be represented in a vector form.
Here, one-dimensional two-body systems are considered as an example,
and their coordinates are denoted by $ x $ and $ y $.
Each direction is discretized with $ M $ meshes,
i.e., in total $ M \times M $ meshes.
Then, the discretized wave function $ \tilde{\psi} $ is
\begin{equation}
  \tilde{\psi}
  =
  \begin{pmatrix}
    \psi_{11} \\
    \psi_{12} \\
    \vdots    \\
    \psi_{1 \left( M - 1 \right)} \\
    \psi_{21} \\
    \psi_{22} \\
    \vdots    \\
    \psi_{2 \left( M - 1 \right)} \\
    \vdots    \\
    \psi_{\left( M - 1 \right) 1} \\
    \psi_{\left( M - 1 \right) 2} \\
    \vdots    \\
    \psi_{\left( M - 1 \right) \left( M - 1 \right)}  
  \end{pmatrix},
\end{equation}
where $ \psi_{jk} = \psi \left( x_j, y_k \right) $,
$ x_j = - x_{\urm{max}} + hj $,
$ y_k = - y_{\urm{max}} + hk $,
and $ x_{\urm{max}} = y_{\urm{max}} $.
Accordingly, the discretized Hamiltonian $ \tilde{H} $ reads
\begin{equation}
  \tilde{H}
  =
  -
  \frac{1}{2 h^2}
  \tilde{T}_1
  -
  \frac{1}{2 h^2}
  \tilde{T}_2
  +
  \tilde{V}_{\urm{ext}}^1
  +
  \tilde{V}_{\urm{ext}}^2
  +
  \tilde{V}_{\urm{int}} ,
\end{equation}
where
$ \tilde{T}_1 $ and $ \tilde{T}_2 $ are the kinetic energy matrices
\begin{subequations}
  \label{eq:2body_matrix_kin}
  \begin{align}
    \tilde{T}_1
    & =
      T \otimes I_{M-1}, \\
    \tilde{T}_2
    & =
      I_{M-1} \otimes T, 
  \end{align}
\end{subequations}
$ \tilde{V}_{\urm{ext}}^1 $ and $ \tilde{V}_{\urm{ext}}^2 $ are the external potential matrices
\begin{subequations}
  \label{eq:2body_matrix_ext}
  \begin{align}
    \tilde{V}^{\urm{ext}}_1
    & =
      V^{\urm{ext}} \otimes I_{M-1}, \\
    \tilde{V}^{\urm{ext}}_2
    & =
      I_{M-1} \otimes V^{\urm{ext}}, 
  \end{align}
\end{subequations}
and $ \tilde{V}_{\urm{int}} $ are the interaction matrix
whose matrix elements are
\begin{widetext}
  \begin{equation}
    \left( \tilde{V}_{\urm{int}} \right)_{i + j \left( M - 1 \right), k + l \left( M - 1 \right)}
    =
    \begin{cases}
      \frac{1}{2}
      \left[
        V^{\urm{int}} \left( x_i, y_j \right)
        +
        V^{\urm{int}} \left( y_j, x_i \right)
      \right]
      =
      V^{\urm{int}} \left( x_i, y_j \right)
      & \text{(for $ i = k $, $ j = l $)}, \\
      0  & \text{(otherwise)},
    \end{cases}
  \end{equation}
\end{widetext}
where $ I_{M-1} $ is the $ \left( M - 1 \right) \times \left( M - 1 \right) $ identity matrix 
\begin{equation}
  I_{M-1}
  =
  \begin{pmatrix}
    1 & 0 & \cdots & 0 \\
    0 & 1 & \cdots & 0 \\
    \vdots & \vdots & \ddots & \vdots \\
    0 & 0 & \cdots & 1 
  \end{pmatrix}
\end{equation}
and $ \otimes$ is the Kronecker product.
For instance, the matrix elements of Eqs.~\eqref{eq:2body_matrix_kin} reads
\newpage
\begin{subequations}
  \begin{align}
    \left( \tilde{T}_1 \right)_{i + j \left( M - 1 \right), k + l \left( M - 1 \right)}
    & =
      \begin{cases}
        -2 & \text{(for $ i = k $, $ j = l $)}, \\
        1  & \text{(for $ i = k $, $ j = l \pm 1 $)}, \\
        0  & \text{(otherwise)},
      \end{cases} \\
    \left( \tilde{T}_2 \right)_{i + j \left( M - 1 \right), k + l \left( M - 1 \right)}
    & =
      \begin{cases}
        -2 & \text{(for $ i = k $, $ j = l $)}, \\
        1  & \text{(for $ i = k \pm 1 $, $ j = l $)}, \\
        0  & \text{(otherwise)}.
      \end{cases} 
  \end{align}
\end{subequations}
\subsubsection{Symmetrization and antisymmetrization}
\label{subsubsec:1d_2body_sym}
\par
The discretized wave function $ \tilde{\psi} $ should be symmetrized or antisymmetrized.
In general, for the arbitrary function $ f \left( x, y \right) $,
$ f \left( x, y \right) + f \left( y, x \right) $
is a symmetrized function and
$ f \left( x, y \right) - f \left( y, x \right) $
is an antisymmetrized function.
\par
To perform (anti)symmetrization in \textsc{Tensorflow},
instead of the simple $ \tilde{\psi} $,
\begin{equation}
  \tilde{\psi}_{\pm}
  =
  \begin{pmatrix}
    \psi_{11} \\
    \psi_{12} \\
    \vdots    \\
    \psi_{1 \left( M - 1 \right)} \\
    \psi_{21} \\
    \psi_{22} \\
    \vdots    \\
    \psi_{2 \left( M - 1 \right)} \\
    \vdots    \\
    \psi_{\left( M - 1 \right) 1} \\
    \psi_{\left( M - 1 \right) 2} \\
    \vdots    \\
    \psi_{\left( M - 1 \right) \left( M - 1 \right)}
  \end{pmatrix}
  \pm
  \begin{pmatrix}
    \psi_{11} \\
    \psi_{21} \\
    \vdots    \\
    \psi_{\left( M - 1 \right) 1} \\
    \psi_{12} \\
    \psi_{22} \\
    \vdots    \\
    \psi_{\left( M - 1 \right) 2} \\
    \vdots    \\
    \psi_{1 \left( M - 1 \right)} \\
    \psi_{2 \left( M - 1 \right)} \\
    \vdots    \\
    \psi_{\left( M - 1 \right) \left( M - 1 \right)}  
  \end{pmatrix}
\end{equation}
is assumed to be a trial wave function.
In the \textsc{Tensorflow} code,
instead of the original \texttt{predicts} and \texttt{output\_wf},
$ \tilde{\psi}_{\pm} $ is used in the third (the calculation process of the loss function) and fifth (calculate the ground-state energy) steps
in Sec.~\ref{subsec:network}.
Note that this process can be easily done by using the following commands:
\begin{enumerate}
\item \texttt{predicts\_transpose = tf.reshape(predicts, [m, m])},
\item \texttt{predicts\_transpose = tf.transpose(predicts\_transpose)},
\item \texttt{predicts\_transpose = tf.reshape(predicts\_transpose, [m**2, 1])},
\item \texttt{predicts = tf.add(predicts, predicts\_transpose)} for bosonic systems
  or
  \texttt{predicts = tf.subtract(predicts, predicts\_transpose)} for fermionic systems,
\item the final \texttt{predicts} is used to evaluate the loss function.
\end{enumerate}
Here, \texttt{m} corresponds to the number of meshes $ M $.
Note that this method can be straightforwardly extended to multibody systems.
\subsubsection{Two-body systems}
\label{subsubsec:1d_2body_results}
\par
Two-body systems under the harmonic oscillator potential [Eq.~\eqref{eq:pot_harmonic}] is tested.
If there is no interaction $ V^{\urm{int}} \equiv 0 $,
the ground-state wave function $ \psi_{\urm{gs}} $ and energy $ E_{\urm{gs}} $ are known exactly.
If one considers bosonic systems,
they read
\begin{subequations}
  \begin{align}
    \psi_{\urm{gs}} \left( x, y \right)
    & =
      \sqrt{\frac{\omega}{\pi}}
      \exp
      \left[
      - \frac{\omega \left( x^2 + y^2 \right)}{2}
      \right],
      \label{eq:2body_ho_boson_wf} \\
    E_{\urm{gs}}
    & =
      \omega,
  \end{align}
\end{subequations}
and if one considers the fermionic systems,
they read
\begin{subequations}
  \begin{align}
    \psi_{\urm{gs}} \left( x, y \right)
    & =
      \frac{\omega}{\sqrt{\pi}}
      \left( x - y \right)
      \exp
      \left[
      - \frac{\omega \left( x^2 + y^2 \right)}{2}
      \right], 
      \label{eq:2body_ho_fermion_wf} \\
    E_{\urm{gs}}
    & =
      2
      \omega.
  \end{align}
\end{subequations}
\par
\begin{figure*}[tb]
  \centering
  \includegraphics[width=1.0\linewidth]{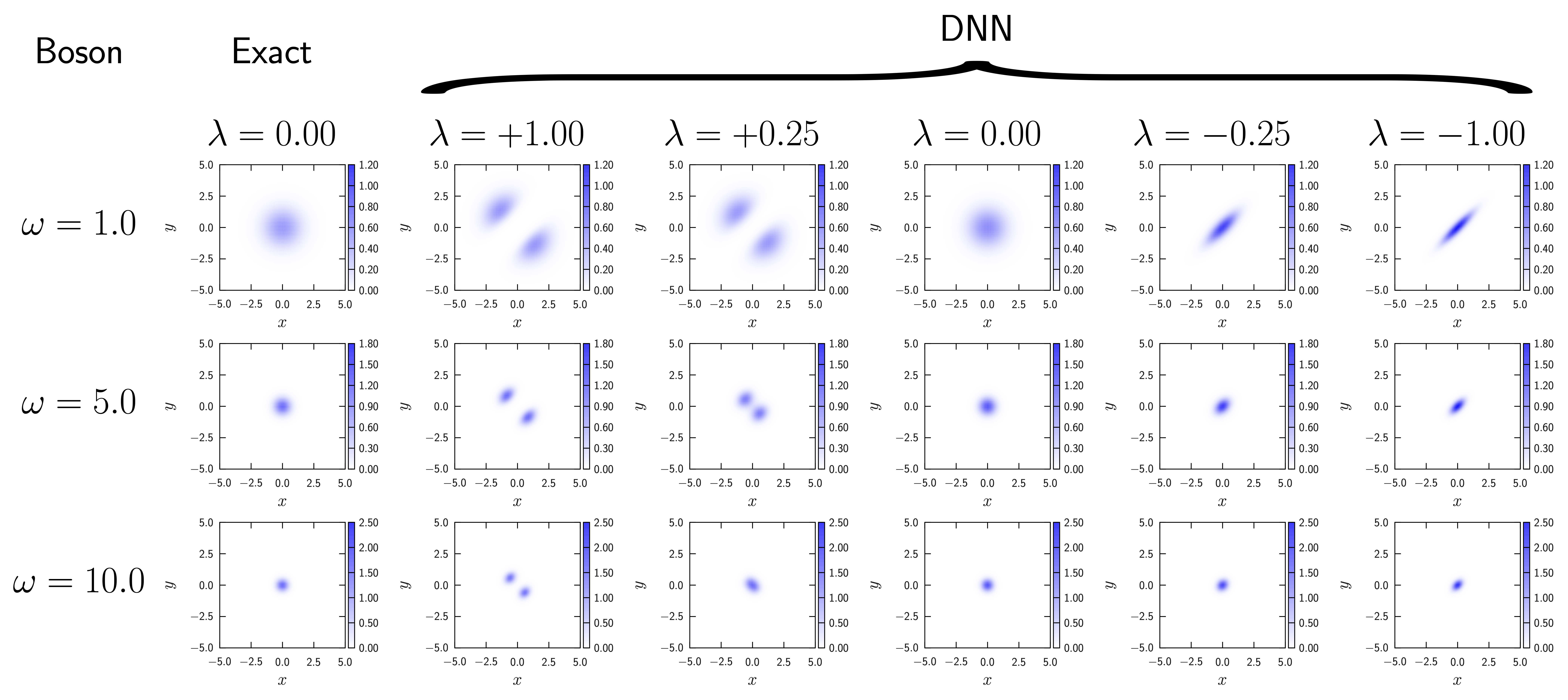}
  \caption{
    Two-body wave function under the harmonic oscillator potential for bosonic systems.
    The exact wave function without the interaction is shown in the left-most column.}
  \label{fig:2body_ho_wf_boson}
\end{figure*}
\begin{figure*}[tb]
  \centering
  \includegraphics[width=1.0\linewidth]{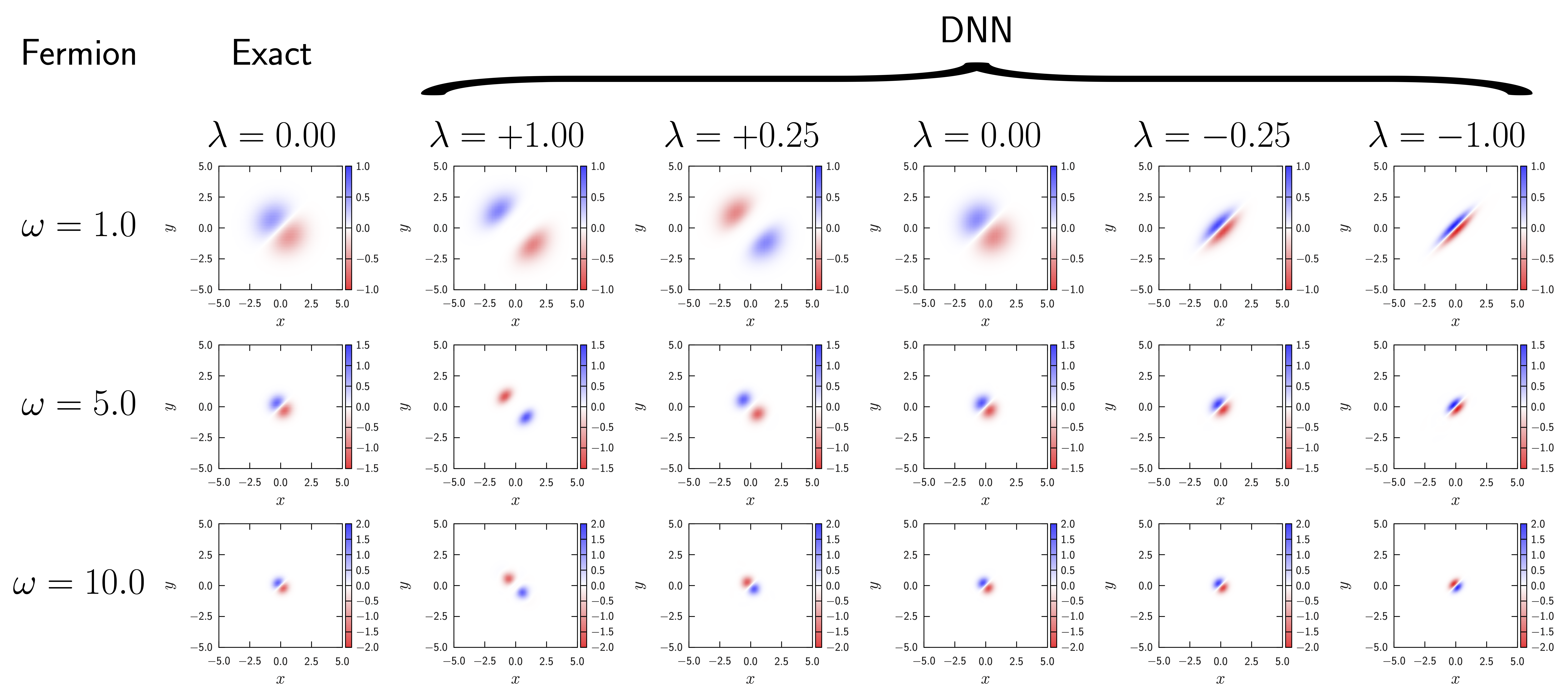}
  \caption{
    Same as Fig.~\ref{fig:2body_ho_wf_boson} but for fermionic systems.}
  \label{fig:2body_ho_wf_fermion}
\end{figure*}

Figures~\ref{fig:2body_ho_wf_boson} and \ref{fig:2body_ho_wf_fermion}, respectively, show wave functions for bosonic and fermionic systems obtained in this work.
The total energies and the calculation time are shown in Table~\ref{tab:2body_ho}.
For comparison, the exact wave functions [Eq.~\eqref{eq:2body_ho_boson_wf} or Eq.~\eqref{eq:2body_ho_fermion_wf}] are also shown.
Here, $ x_{\urm{max}} = y_{\urm{max}} = 5 $ and $ M_x = M_y = 256 $ are used for the spatial mesh
and
two layers each of which contains $ 32 $ units are used for the DNN.
For the interaction, the Gaussian-type interaction
\begin{equation}
  \label{eq:interaction}
  V^{\urm{int}} \left( x, y \right)
  =
  100 \lambda
  \exp \left( - \left( x - y \right)^2 \right)
\end{equation}
is used, where $ \lambda $ is the strength of the interaction.
The DNN results with $ \lambda = 0 $ show a good agreement with the exact results,
demonstrating that the DNN technique works well.
\par
Behavior of wave functions for nonzero $ \lambda $ 
is consistent qualitatively with our expectation:
if $ \lambda $ is negative, i.e., the interaction is attractive, the wave function tends to collapse,
and
if $ \lambda $ is positive, i.e., the interaction is repulsive, the wave function tends to be broad.
Time cost per epoch is almost universal among all the calculation,
while more epochs are required to reach convergence for fermionic systems than for bosonic systems.
This may be because all the values are positive for the initial condition,
while there are negative values for fermionic ground-state wave functions.
More epochs are required for the repulsive interaction ($ \lambda > 0 $) than for the attractive interaction ($ \lambda < 0 $).
This may be because the topology of the wave function is more complicated and extended in the repulsive case than the attractive case.
\par
\begin{figure*}[tb]
  \centering
  \includegraphics[width=1.0\linewidth]{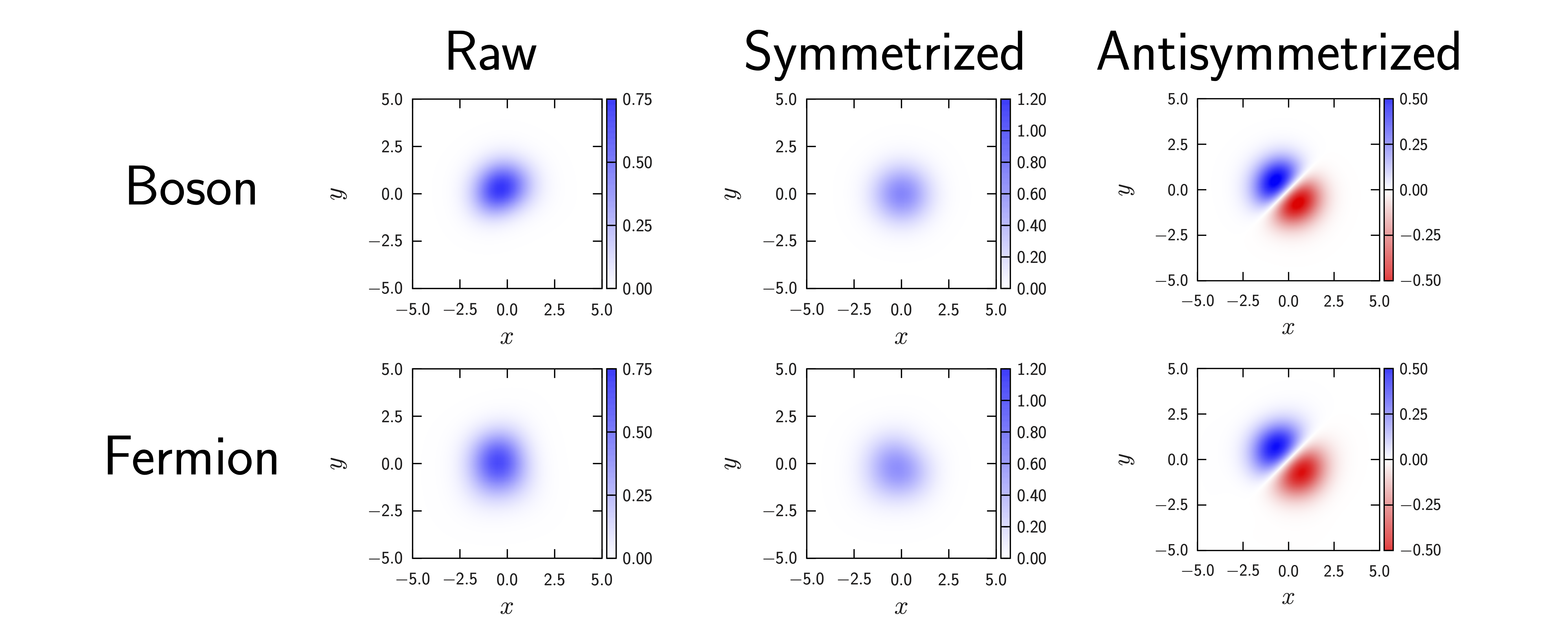}
  \caption{
    DNN wave function of the raw ($ u_{\urm{gs}} $ in Eq.~\eqref{eq:softplus_2body}), the symmetrized, and the antisymmetrized wave functions.
    The rows named ``Boson'' and ``Fermion,'' respectively, correspond to the results obtained by minimizing
    the bosonic or fermionic energy expectation value.}
  \label{fig:2body_analysis}
\end{figure*}

Finally, we point out a strange behavior of the obtained wave function of the two-body system of $ \omega = 1.0 $ without the interaction.
Here, for simplicity, the two-layer DNN in which each layer is composed of four units is used.
In the case of two layers, 
the function obtained by optimized weights of DNN is 
\begin{subequations}
  \begin{align}
    u_{\urm{gs}} \left( x, y \right)
    & =
      A
      \softplus \left( \sum_{j = 1}^{N_{\uurm{unit}}} w_{2j} u_{2j} \left( x, y \right) + b_2 \right),
      \label{eq:softplus_2body} \\
    u_{2j} \left( x , y \right)
    & =
      \softplus \left( \sum_{k = 1}^{N_{\uurm{unit}}} w_{1jk} u_{1k} \left( x, y \right) + b_{1j} \right), \\
    u_{1k} \left( x, y \right)
    & =
      \softplus \left( w_{0k0} x + w_{0k1} y + b_{0k} \right),
  \end{align}
\end{subequations}
where
$ A $ is the normalization constant,
$ N_{\urm{unit}} $ is the number of units of each layer,
$ w $ is a weight, and $ b $ is a bias.
The left column of Fig.~\ref{fig:2body_analysis} shows $ u_{\urm{gs}} \left( x, y \right) $.
The upper and lower rows, respectively, correspond to the results with minimizing the bosonic or fermionic energy expectation value.
It is shown that the obtained function $ u_{\urm{gs}} $,
which is referred to as the raw \textit{wave function},
is not symmetric nor antisymmetric.
After the symmetrization
$ \psi_{\urm{sym}} \left( x, y \right) = \left[ u_{\urm{gs}} \left( x, y \right) + u_{\urm{gs}} \left( y, x \right) \right] / A_{\urm{sym}} $
or
the antisymmetrization
$ \psi_{\urm{antisym}} \left( x, y \right) = \left[ u_{\urm{gs}} \left( x, y \right) - u_{\urm{gs}} \left( y, x \right) \right] / A_{\urm{antisym}} $
is performed
with the normalization constant $ A_{\urm{sym}} $ or $ A_{\urm{antisym}} $,
$ \psi_{\urm{sym}} $ or $ \psi_{\urm{antisym}} $
can be regarded as 
the bosonic or fermionic ground-state wave function, respectively.
The energy expectation value of the raw ($ u_{\urm{gs}} $),
the symmetrized ($ \psi_{\urm{sym}} $),
and the antisymmetrized ($ \psi_{\urm{antisym
  }} $) wave functions are shown in Table~\ref{tab:2body_dnn}.
A surprising fact is that even if the raw \textit{wave function} is obtained by minimizing the bosonic (fermionic) expectation value,
fermionic (bosonic) energy expectation value is close to the correct fermionic (bosonic) energy eigenvalue,
and \textit{vice versa}.
The (anti)symmetrization corresponds to the projection of the raw \textit{wave function}
to the boson (fermion) subspace,
while the remaining part is not supposed to be optimized well.
This unexpected coincidence may be due to the smallness of the number of parameters in the DNN architecture,
and deserves further study.
\begin{table*}[tb]
  \centering
  \caption{
    Calculation summary of a two-body problem under the harmonic oscillator potential.}
  \label{tab:2body_ho}
  \begin{ruledtabular}
    \begin{tabular}{lddddd}
      \multicolumn{1}{c}{Particles} & \multicolumn{1}{c}{$ \omega $} & \multicolumn{1}{c}{$ \lambda $} & \multicolumn{1}{c}{Energy} & \multicolumn{1}{c}{\# of Epochs} & \multicolumn{1}{c}{Time per Epoch ($ \mathrm{ms} $)} \\
      \hline
      Boson & 1.0 & -1.00 & -89.869381 & 8490 & 23.583 \\
      Boson & 1.0 & -0.25 & -19.848949 & 3595 & 23.618 \\
      Boson & 1.0 & +0.00 & +0.999927 & 27242 & 23.424 \\
      Boson & 1.0 & +0.25 & +3.298725 & 20040 & 23.529 \\
      Boson & 1.0 & +1.00 & +3.835173 & 21763 & 23.677 \\
      \hline
      Boson & 5.0 & -1.00 & -87.554311 & 10194 & 23.573 \\
      Boson & 5.0 & -0.25 & -17.203647 & 10893 & 23.712 \\
      Boson & 5.0 & +0.00 & +4.997829 & 24477 & 23.648 \\
      Boson & 5.0 & +0.25 & +21.149827 & 20721 & 23.797 \\
      Boson & 5.0 & +1.00 & +31.804917 & 21718 & 23.591 \\
      \hline
      Boson & 10.0 & -1.00 & -84.129658 & 18635 & 23.566 \\
      Boson & 10.0 & -0.25 & -13.118213 & 24380 & 23.601 \\
      Boson & 10.0 & +0.00 & +9.991009 & 27489 & 23.692 \\
      Boson & 10.0 & +0.25 & +32.287424 & 23810 & 23.816 \\
      Boson & 10.0 & +1.00 & +72.688350 & 19591 & 23.601 \\
      \hline
      Fermion & 1.0 & -1.00 & -71.409493 & 18794 & 23.928 \\
      Fermion & 1.0 & -0.25 & -11.369632 & 17999 & 23.818 \\
      Fermion & 1.0 & +0.00 & +1.999931 & 19215 & 23.843 \\
      Fermion & 1.0 & +0.25 & +3.298786 & 25187 & 23.804 \\
      Fermion & 1.0 & +1.00 & +3.839178 & 106163 & 24.136 \\
      \hline
      Fermion & 5.0 & -1.00 & -68.409494 & 18558 & 23.731 \\
      Fermion & 5.0 & -0.25 & -7.207718 & 19618 & 23.787 \\
      Fermion & 5.0 & +0.00 & +9.995902 & 75208 & 23.841 \\
      Fermion & 5.0 & +0.25 & +21.385884 & 51691 & 23.607 \\
      Fermion & 5.0 & +1.00 & +31.804915 & 27885 & 23.799 \\
      \hline
      Fermion & 10.0 & -1.00 & -63.026667 & 26603 & 23.894 \\
      Fermion & 10.0 & -0.25 & +0.302312 & 21783 & 23.805 \\
      Fermion & 10.0 & +0.00 & +19.975414 & 10135 & 23.741 \\
      Fermion & 10.0 & +0.25 & +38.060907 & 79973 & 24.054 \\
      Fermion & 10.0 & +1.00 & +73.245097 & 59289 & 23.972 \\
    \end{tabular}
  \end{ruledtabular}
\end{table*}
\begin{table}[tb]
  \centering
  \caption{
    Energy expectation value of the raw ($ u_{\urm{gs}} $ in Eq.~\eqref{eq:softplus_2body}), the symmetrized, and the antisymmetrized wave functions.
    The rows named ``Boson'' and ``Fermion,'' respectively, correspond to the results obtained by minimizing
    the bosonic or fermionic energy expectation value.}
  \label{tab:2body_dnn}
  \begin{ruledtabular}
    \begin{tabular}{lddd}
      & \multicolumn{1}{c}{Raw} & \multicolumn{1}{c}{Symmetrized} & \multicolumn{1}{c}{Antisymmetrized} \\
      \hline
      Boson   & 1.13190 & 1.00012 & 2.05195 \\
      Fermion & 1.13628 & 1.08235 & 2.00056 \\
    \end{tabular}
  \end{ruledtabular}
\end{table}
\subsubsection{Three-body systems}
\label{subsubsec:1d_3body_results}
\par
Three-body systems under the harmonic oscillator potential [Eq.~\eqref{eq:pot_harmonic}] is tested.
For simplicity, we consider a system without any interaction $ V^{\urm{int}} \equiv 0 $.
Then, the ground-state wave function $ \psi_{\urm{gs}} $ and energy $ E_{\urm{gs}} $ are known exactly as 
\begin{subequations}
  \begin{align}
    \psi_{\urm{gs}} \left( x, y, z \right)
    & =
      \left(
      \frac{\omega}{\pi}
      \right)^{3/4}
      \exp
      \left[
      - \frac{\omega \left( x^2 + y^2 + z^2 \right)}{2}
      \right], 
      \label{eq:3body_ho_boson_wf} \\
    E_{\urm{gs}}
    & =
      \frac{3}{2}
      \omega,
  \end{align}
\end{subequations}
for bosonic systems
and
\begin{widetext}
  \begin{subequations}
    \begin{align}
      \psi_{\urm{gs}} \left( x, y, z \right)
      & =
        \left(
        \frac{\omega}{\pi}
        \right)^{3/4}
        \sqrt{\frac{\omega}{6}}
        \left[
        \left( x - y \right)
        \left( 1 - 2 \omega z^2 \right)
        +
        \left( y - z \right)
        \left( 1 - 2 \omega x^2 \right)
        +
        \left( z - x \right)
        \left( 1 - 2 \omega y^2 \right)
        \right]
        \exp
        \left[
        - \frac{\omega \left( x^2 + y^2 + z^2 \right)}{2}
        \right], 
        \label{eq:3body_ho_fermion_wf} \\
      E_{\urm{gs}}
      & =
        \frac{9}{2}
        \omega
    \end{align}
  \end{subequations}
\end{widetext}
for fermionic systems.
\par
\begin{figure}[tb]
  \centering
  \begin{minipage}{0.49\linewidth}
    \centering
    \includegraphics[width=1.0\linewidth]{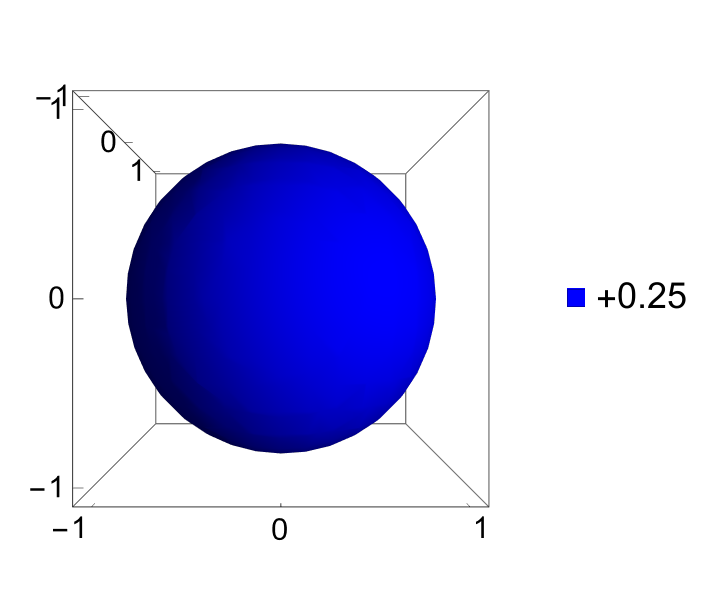}
  \end{minipage}
  \hfill
  \begin{minipage}{0.49\linewidth}
    \centering
    \includegraphics[width=1.0\linewidth]{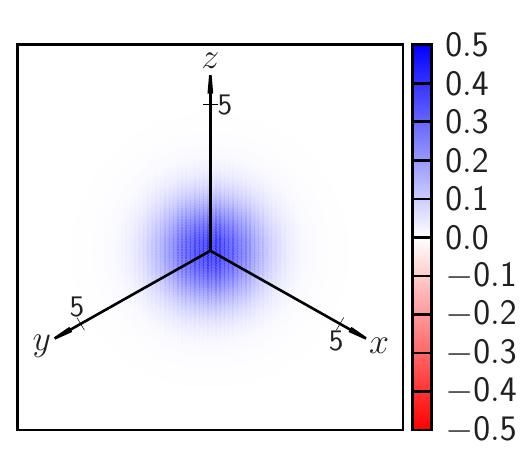}
  \end{minipage}
  \caption{
    (Left) Three-body wave function under the harmonic oscillator potential without interparticle interaction for bosonic systems.
    (Right) Slice of the three-body wave function at the plane $ x + y + z = 0 $.}
  \label{fig:3body_ho_wf_boson}
\end{figure}
\begin{figure}[tb]
  \centering
  \begin{minipage}{0.49\linewidth}
    \centering
    \includegraphics[width=1.0\linewidth]{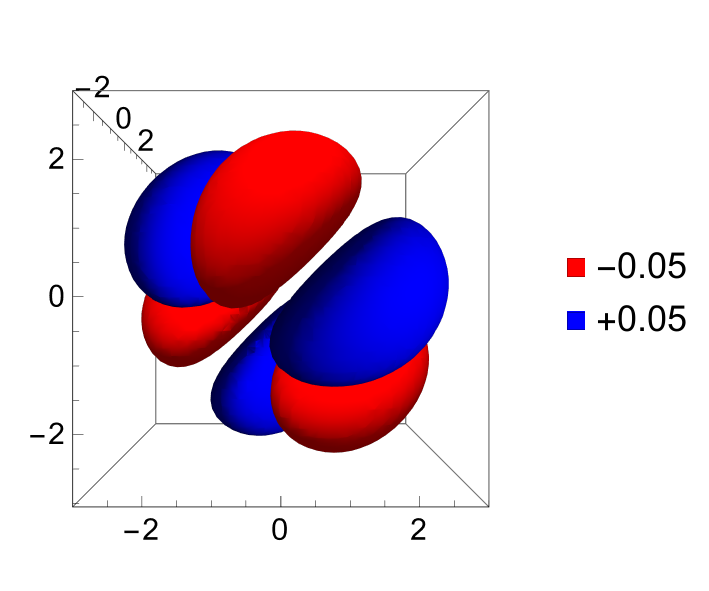}
  \end{minipage}
  \hfill
  \begin{minipage}{0.49\linewidth}
    \centering
    \includegraphics[width=1.0\linewidth]{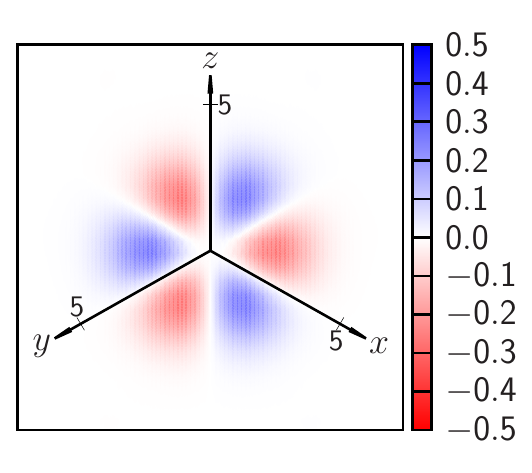}
  \end{minipage}
  \caption{
    Same as Fig.~\ref{fig:3body_ho_wf_boson} but for fermionic systems.}
  \label{fig:3body_ho_wf_fermion}
\end{figure}

Figures~\ref{fig:3body_ho_wf_boson} and \ref{fig:3body_ho_wf_fermion}, respectively, show wave functions for bosonic and fermionic systems obtained by this work.
The total energies and the calculation time are shown in Table~\ref{tab:3body_ho}.
Here, $ x_{\urm{max}} = y_{\urm{max}} = z_{\urm{max}} = 5 $ and $ M_x = M_y = M_z = 64 $ are used for the spatial mesh
and
two layers each of which contains $ 32 $ units are used for the DNN.
The interaction is not considered.
\par
The DNN calculations reproduce the exact ground-state energies.
The DNN wave functions are consistent with the exact solution.
The number of epochs for three-body systems is comparable with those for two-body systems,
where the number of units and layers are identical for these two cases.
In contrast, the time per epoch for the three-body systems are about four times of that for the two-body systems.
This is related to the number of spatial meshes:
$ 256 \times 256 = 65536 $ meshes are used for the two-body systems
and
$ 64 \times 64 \times 64 = 262144 $ meshes are used for the three-body systems;
thus, the number of meshes for the three-body systems are four times more than those for the two-body systems.
Hence, it can be concluded that the time per epoch is almost proportional to the number of spatial meshes.
This is reasonable since we use numerical methods for sparse matrices,
in which the number of the nonzero matrix elements is $ O \left( M^{Nd} \right) $.
\begin{table}[tb]
  \centering
  \caption{
    Calculation summary of a three-body problem under the harmonic oscillator potential.
    Calculation is performed with $ \omega = 1.0 $.}
  \label{tab:3body_ho}
  \begin{ruledtabular}
    \begin{tabular}{lddd}
      \multicolumn{1}{c}{Particles} & \multicolumn{1}{c}{Energy} & \multicolumn{1}{c}{\# of Epochs} & \multicolumn{1}{c}{Time per Epoch ($ \mathrm{ms} $)} \\
      \hline
      Boson & +1.497880 & 20183 & 101.216 \\
      Fermion & +4.486830 & 22770 & 98.356 \\
    \end{tabular}
  \end{ruledtabular}
\end{table}
%
% Excited State
%
\section{Excited-State Calculation}
\label{sec:excited_state}
\par
In this section, based on the variational principle, a method to calculate low-lying excited states sequentially is explained.
Assume that wave functions of the ground state and $ n $ excited states,
$ \ket{\psi_0} $, $ \ket{\psi_1} $, \ldots, $ \ket{\psi_n} $,
are obtained, where $ \ket{\psi_0} = \ket{\psi_{\urm{gs}}} $.
We consider a problem of finding 
the $ \left( n + 1 \right) $-th excited state $ \ket{\psi_{n + 1}} $,
which satisfies the orthnormal condition
\begin{equation}
  \braket{\psi_j}{\psi_{n + 1}}
  =
  \delta_{j, n + 1} ,
\end{equation}
by using a trial wave function $ \ket{\psi} $.
The $ \left( n + 1 \right) $-th wave function can be obtained
with minimizing the expectation value
\begin{equation}
  \avr{H}
  =
  \frac{\brakket{\psi}{H}{\psi}}{\braket{\psi}{\psi}},
\end{equation}
where $ \ket{\psi} $ is assumed to be orthogonal to $ \ket{\psi_j} $ ($ j = 0 $, $ 1 $, \ldots, $ n $).
This can be implemented in \textsc{Tensorflow}
with assuming that 
\begin{equation}
  \label{eq:orthogonal}
  \ket{\psi}
  -
  \sum_{j = 0}^n
  \braket{\psi_j}{\psi}
  \ket{\psi_j}
\end{equation}
is a trial wave function,
instead of the simple $ \ket{\psi} $.
For one-body problem, $ x_{\urm{max}} = 5 $ and $ M = 1024 $ are used for the spatial mesh
and
the single-layer DNN with eight units is adopted.
\subsection{Harmonic oscillator potential}
\label{eq:subsec_ex_ho}
\par
One-body one-dimensional harmonic oscillators are taken as examples.
The exact wave functions for several low-lying excited states are~\cite{
  Schiff1968QuantumMechanics_McGrawHill}
\begin{subequations}
  \label{eq:ho_ex}
  \begin{align}
    \psi_0 \left( x \right)
    & =
      \left( \frac{\omega}{\pi} \right)^{1/4}
      \exp
      \left( - \frac{\omega x^2}{2} \right), \\
    \psi_1 \left( x \right)
    & =
      \left( \frac{\omega}{\pi} \right)^{1/4}
      \sqrt{2 \omega} x 
      \exp
      \left( - \frac{\omega x^2}{2} \right), \\
    \psi_2 \left( x \right)
    & =
      \left( \frac{\omega}{\pi} \right)^{1/4}
      \frac{2 \omega x^2 - 1}{\sqrt{2}}
      \exp
      \left( - \frac{\omega x^2}{2} \right), \\
    \psi_3 \left( x \right)
    & =
      \left( \frac{\omega}{\pi} \right)^{1/4}
      \sqrt{\frac{\omega}{3}}
      \left( 2 \omega x^2 - 3 \right) x 
      \exp
      \left( - \frac{\omega x^2}{2} \right), \\
    \psi_4 \left( x \right)
    & =
      \left( \frac{\omega}{\pi} \right)^{1/4}
      \frac{4 \omega^2 x^4 - 12 \omega x^2 + 3}{2 \sqrt{6}}
      \exp
      \left( - \frac{\omega x^2}{2} \right),
  \end{align}
\end{subequations}
where $ \psi_n $ is the $ n $-th excited state,
and the energies are
\begin{equation}
  E_n
  =
  \left( n + \frac{1}{2} \right)
  \omega.
\end{equation}
\par
\begin{figure}[tb]
  \centering
  \includegraphics[width=1.0\linewidth]{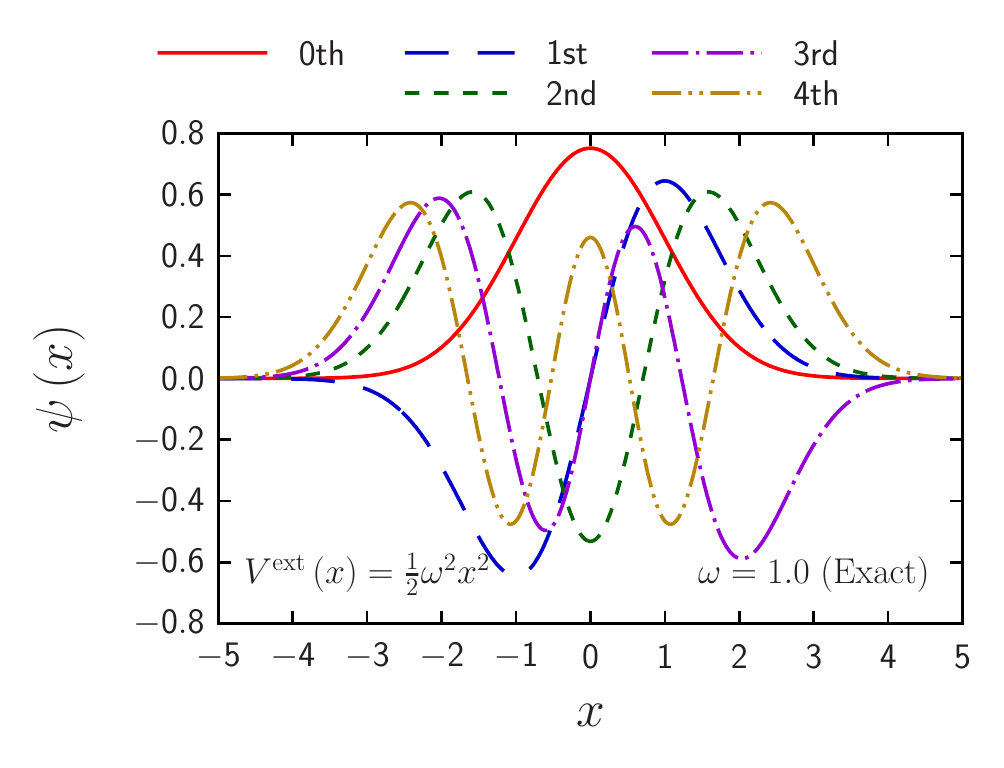}
  \includegraphics[width=1.0\linewidth]{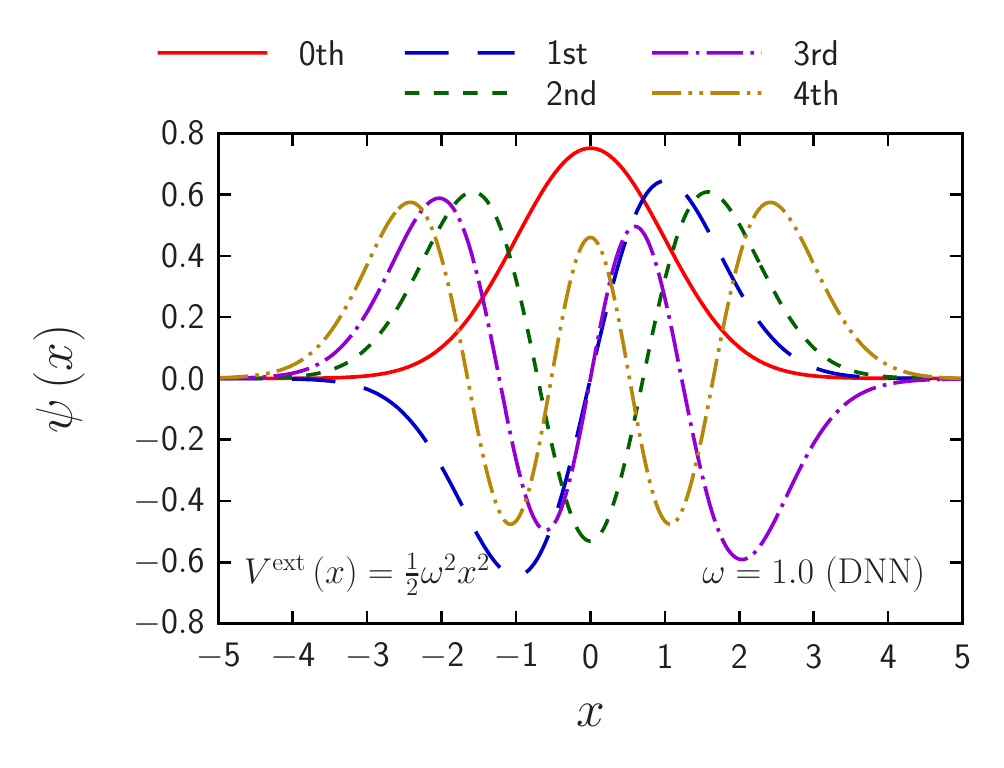}
  \caption{
    Wave functions of the ground and low-lying excited states
    under the harmonic oscillator potential.
    The top panel shows the exact wave functions and
    the bottom one shows the DNN wave functions.
    To make consistency for the phase factor,
    $ -\psi_3 \left( x \right) $ is plotted for the exact wave function of the third excited state.}
  \label{fig:ex_ho_wf}
\end{figure}

Figure~\ref{fig:ex_ho_wf} shows the wave functions of the ground state and
first, second, third, and fourth excited states.
Table~\ref{tab:ex_ho} shows the summary of calculations.
Basically, not only the ground-state but also low-lying excited-states wave functions and energies are successfully calculated.
Thus, it can be concluded that the method to calculate low-lying excited states proposed here works well.
\par
The number of epochs are almost universal for all states calculated here.
In contrast, the time per epoch for a higher excited state is slightly longer
since calculation for orthogonal condition [Eq.~\eqref{eq:orthogonal}] is needed to be performed,
while it takes just a few $ \mathrm{\mu s} $.
\par
\begin{figure}[tb]
  \centering
  \includegraphics[width=1.0\linewidth]{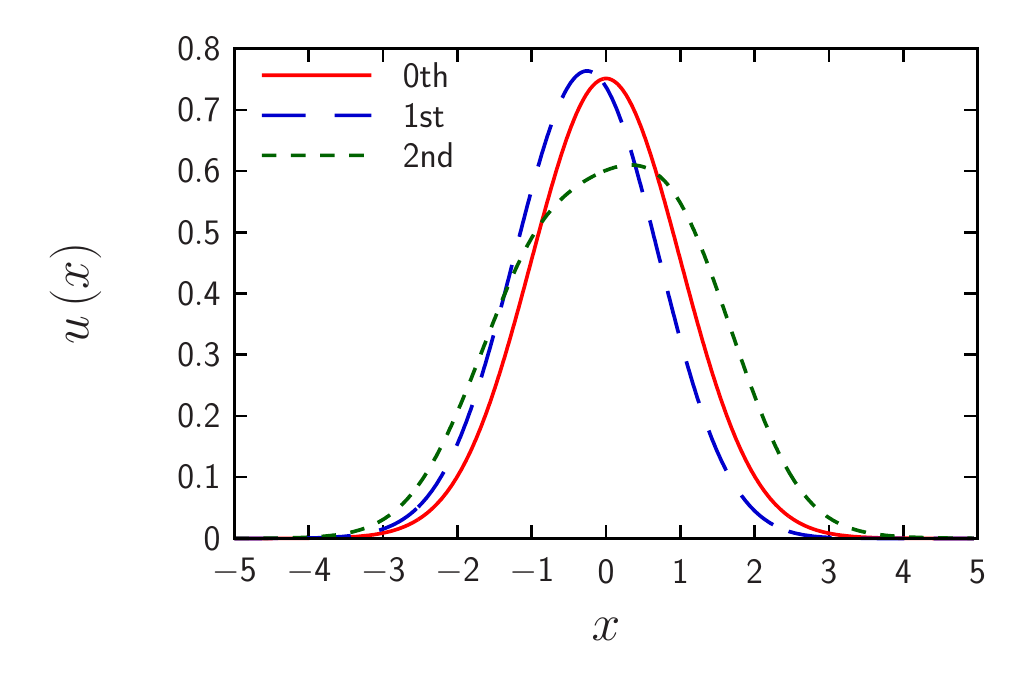}
  \caption{
    Optimized function $ u $ obtained by the DNN for the ground-state (0th),
    the first and the second excited states.}
  \label{fig:dnn_analysis_2d_ho}
\end{figure}

Let us explain why our simple DNN can describe even the excited states correctly.
For simplicity, the single-layer DNN with the four unit is used.
The optimized raw \textit{wave function} for the ground state ($ u_{\urm{gs}} $)
and
the first ($ u_{\urm{1st}} $),
and the second ($ u_{\urm{2nd}} $)
excited states are shown in Fig.~\ref{fig:dnn_analysis_2d_ho}.
The obtained function for the $ n $-th excited state is
\begin{equation}
  u_{\urm{$ n $-th}} \left( x \right)
  =
  \sum_{j = 0}^n
  a_j
  \psi_{\urm{$ n $-th}} \left( x \right)
\end{equation}
with $ \sum_{j = 0}^n \left| a_j \right|^2 = 1 $,
where $ \psi_{\urm{$ n $-th}} $ is the $ n $-th excited-state wave function.
In the case of the first and second excited states,
\begin{widetext}
  \begin{subequations}
    \begin{align}
      u_{\urm{1st}} \left( x \right)
      & =
        0.981758
        \psi_{\urm{0th}} \left( x \right)
        +
        0.190136
        \psi_{\urm{1st}} \left( x \right), \\
      u_{\urm{2nd}} \left( x \right)
      & =
        0.969707
        \psi_{\urm{0th}} \left( x \right)
        -
        0.0445505
        \psi_{\urm{1st}} \left( x \right)
        +
        0.240173
        \psi_{\urm{2nd}} \left( x \right)
    \end{align}
  \end{subequations}
\end{widetext}
are obtained.
We notice that 
the most part of the obtained raw \textit{wave function} is the ground state and the small fraction is for the excited components;
hence, the fairly simple raw \textit{wave function},
which is made by the small architecture of the DNN and 
is close to that of the ground-state wave function,
is capable of describing even the excited states.
\begin{table}[tb]
  \centering
  \caption{
    Calculation summary of excited states for a one-body problem under the harmonic oscillator potential.
    Calculation is performed with $ \omega = 1.0 $.}
  \label{tab:ex_ho}
  \begin{ruledtabular}
    \begin{tabular}{lddd}
      \multicolumn{1}{c}{State} & \multicolumn{1}{c}{Energy} & \multicolumn{1}{c}{Epochs} & \multicolumn{1}{c}{Time per Epoch ($ \mathrm{\mu s} $)} \\
      \hline
      0th & +0.499998 & 23419 & 516.238 \\
      1st & +1.499991 & 25646 & 519.026 \\
      2nd & +2.499986 & 23157 & 527.849 \\
      3rd & +3.500193 & 37880 & 534.115 \\
      4th & +4.500201 & 19101 & 542.224 \\
    \end{tabular}
  \end{ruledtabular}
\end{table}
\subsection{Double-well potential}
\label{eq:subsec_ex_dw}
\par
To see the effect of degeneracy,
we also test the double-well potential
\begin{equation}
  V^{\urm{ext}} \left( x \right)
  =
  \left( x^2 - \alpha^2 \right)^2.
\end{equation}
If the central barrier is low enough, i.e., $ \alpha $ is small enough,
each state is not degenerate.
In contrast, if the central barrier is high, i.e., $ \alpha $ is large,
low-lying excited states below the central barrier are twofold degenerate:
One state $ \psi_{\urm{L}} $ is localized into the left ($ x < 0 $) region
while the other state $ \psi_{\urm{R}} $ is localized into the right ($ x > 0 $) region,
and $ \psi_{\urm{L}} \left( x \right) = \psi_{\urm{R}} \left( - x \right) $ holds.
Using a linear combination of these two degenerate states,
one can recognize each state is degenerate of the following two states:
$ \psi_{\pm} \left( x \right) = \left[ \psi_{\urm{L}} \left( x \right) \pm \psi_{\urm{R}} \left( x \right) \right] / \sqrt{2} $,
where $ \psi_{+} $ ($ \psi_{-} $) is a positive (negative) parity state.
According to the exact diagonalization,
$ \alpha = 1.0 $ and $ 1.25 $ give nondegenerate ground and low-lying excited states
and thus $ \psi_j $ is just a $ j $-th excited state,
while $ \alpha = 2.0 $ and $ 3.0 $ give ground and low-lying excited states,
which are almost twofold degenerate:
$ \psi_0 $ and $ \psi_1 $ correspond to the ground states
and 
$ \psi_2 $ and $ \psi_3 $ correspond to the first excited states.
\par
\begin{figure*}[tb]
  \centering
  \begin{minipage}{0.49\linewidth}
    \centering
    \includegraphics[width=1.0\linewidth]{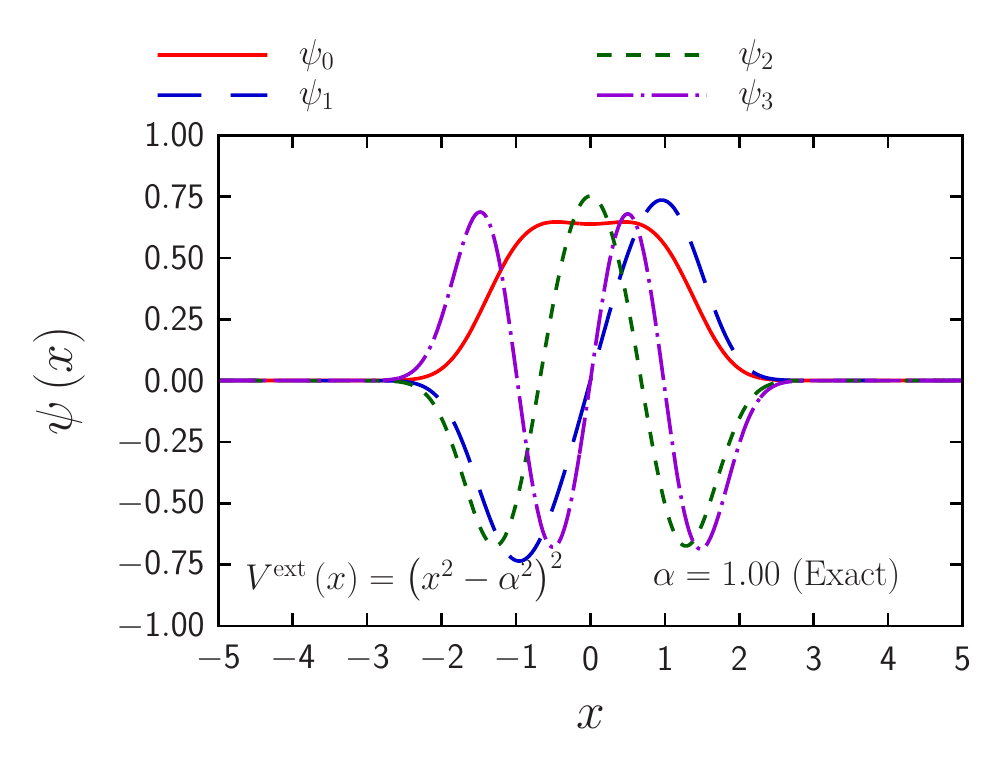}
  \end{minipage}
  \hfill
  \begin{minipage}{0.49\linewidth}
    \centering
    \includegraphics[width=1.0\linewidth]{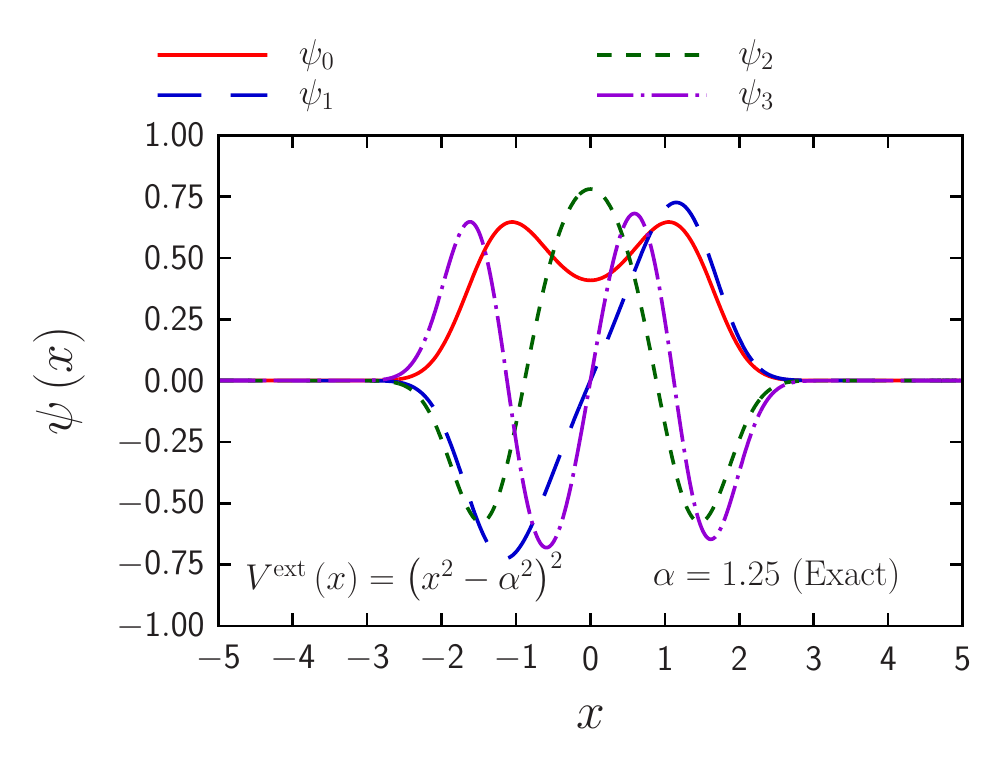}
  \end{minipage}
  \begin{minipage}{0.49\linewidth}
    \centering
    \includegraphics[width=1.0\linewidth]{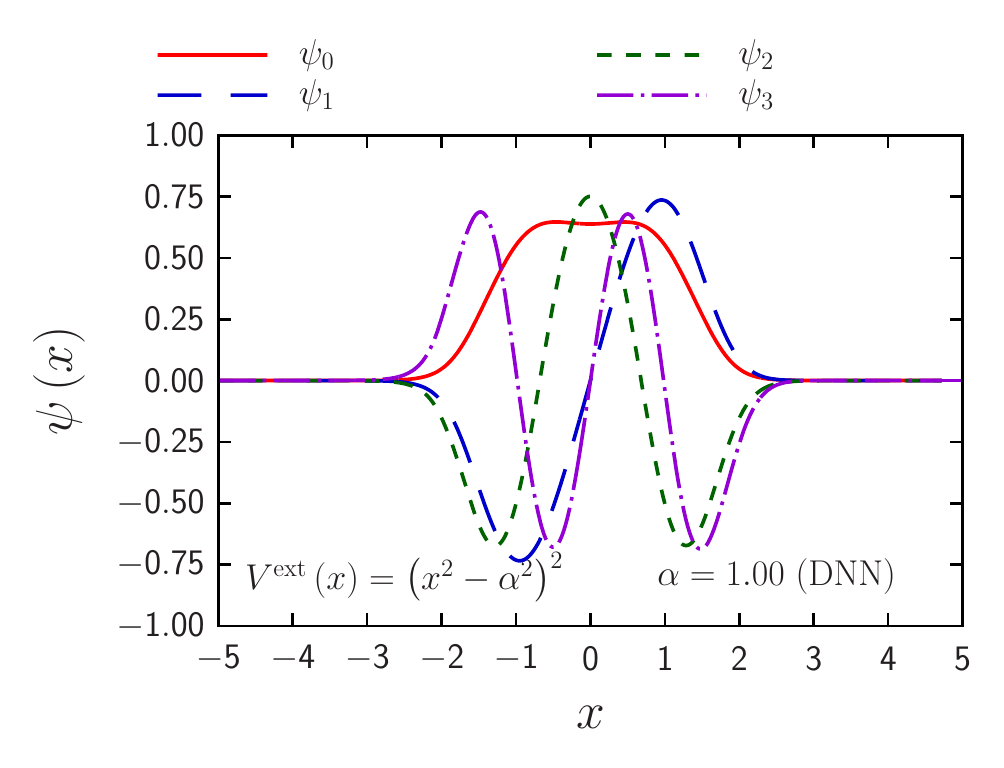}
  \end{minipage}
  \hfill
  \begin{minipage}{0.49\linewidth}
    \centering
    \includegraphics[width=1.0\linewidth]{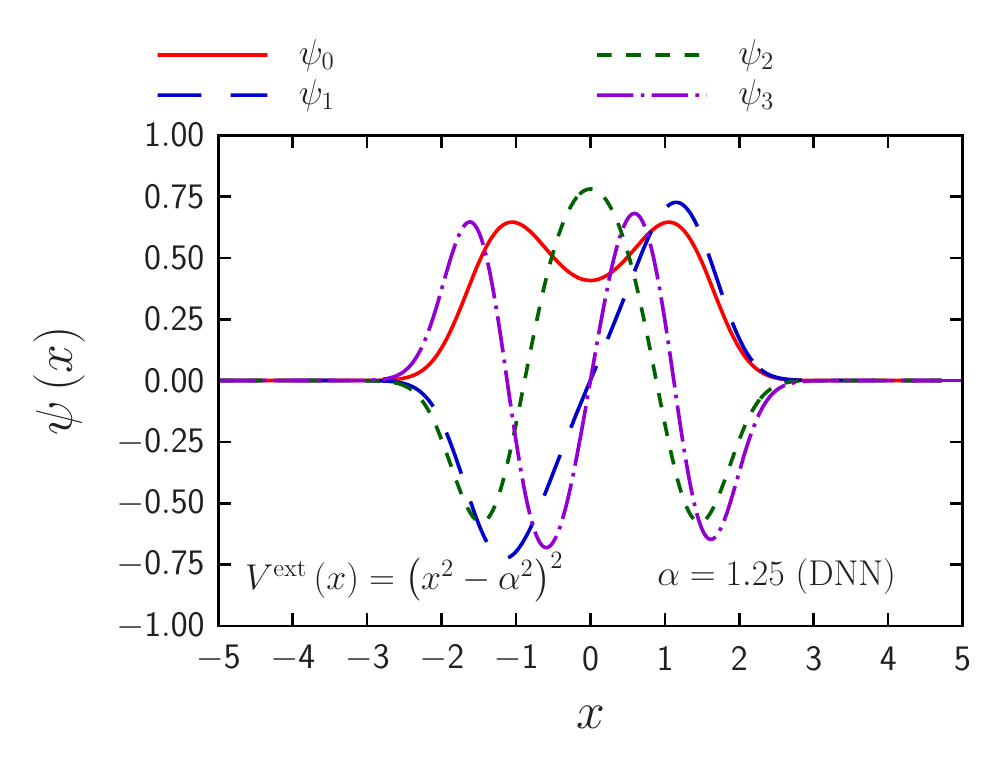}
  \end{minipage}
  \caption{
    Wave functions of the ground and low-lying excited states
    under the double-well potential for $ \alpha = 1.0 $ and $ 1.25 $.
    The top panels show the exact wave functions and
    the bottom ones show the DNN wave functions.}
  \label{fig:ex_dw_wf_1}
\end{figure*}
\begin{figure*}[tb]
  \centering
  \begin{minipage}{0.49\linewidth}
    \centering
    \includegraphics[width=1.0\linewidth]{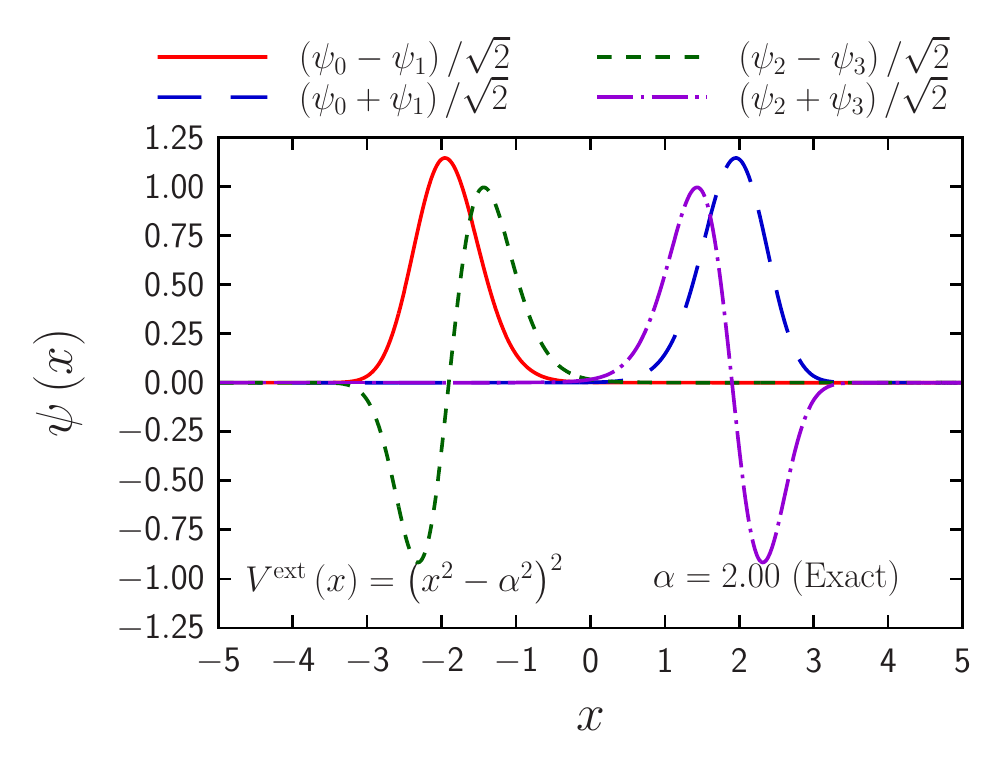}
  \end{minipage}
  \hfill
  \begin{minipage}{0.49\linewidth}
    \centering
    \includegraphics[width=1.0\linewidth]{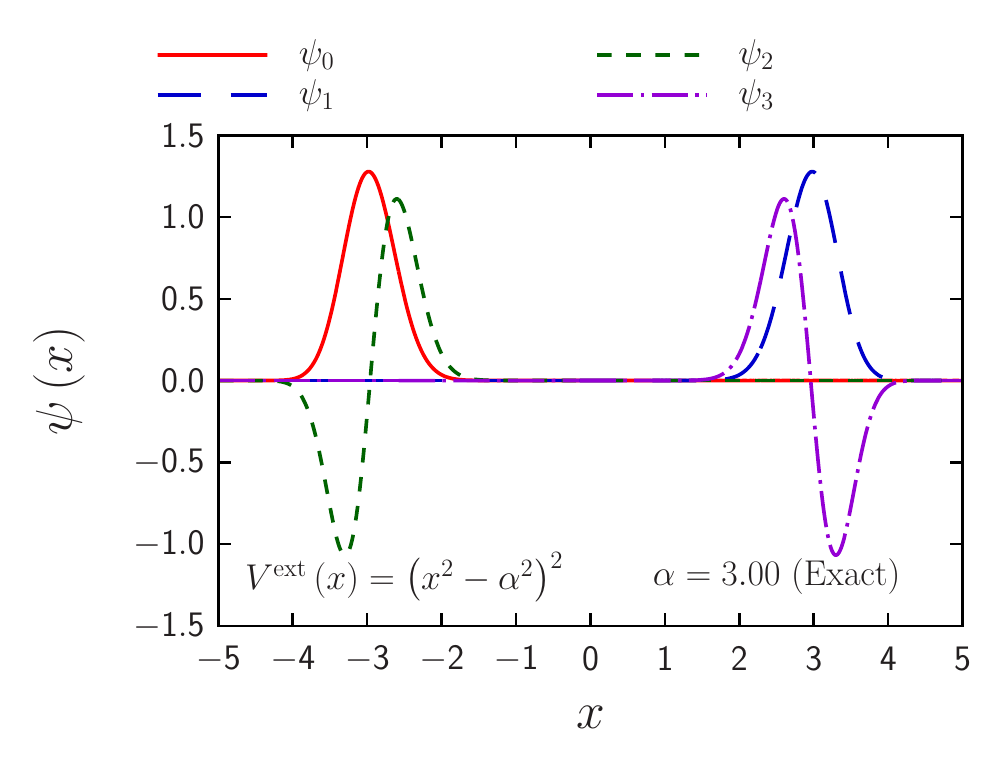}
  \end{minipage}
  \begin{minipage}{0.49\linewidth}
    \centering
    \includegraphics[width=1.0\linewidth]{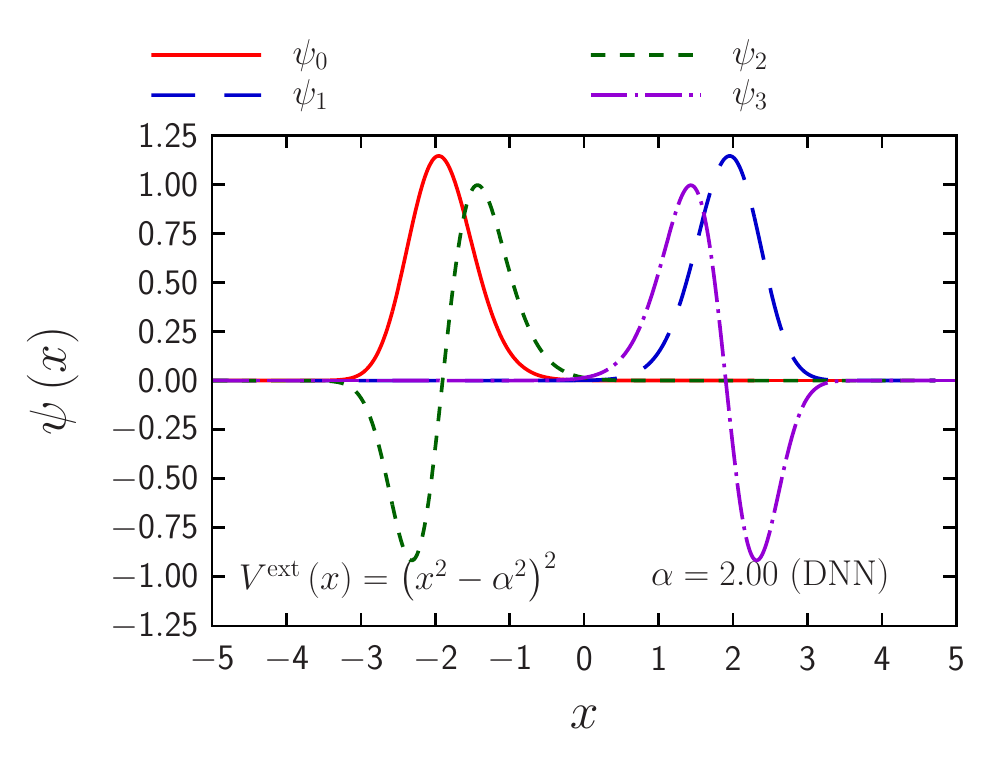}
  \end{minipage}
  \hfill
  \begin{minipage}{0.49\linewidth}
    \centering
    \includegraphics[width=1.0\linewidth]{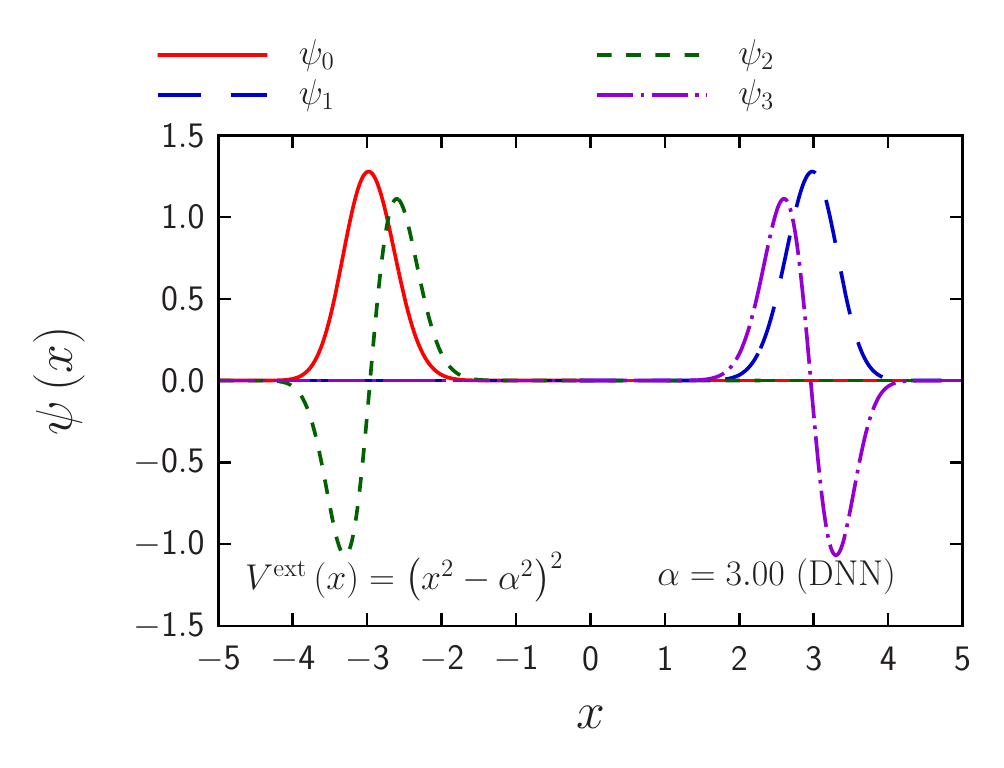}
  \end{minipage}
  \caption{
    Same as Fig.~\ref{fig:ex_dw_wf_1} but for $ \alpha = 2.0 $ and $ 3.0 $.}
  \label{fig:ex_dw_wf_2}
\end{figure*}

Figures~\ref{fig:ex_dw_wf_1} and \ref{fig:ex_dw_wf_2} shows the wave functions of the ground state and
first, second, third, and fourth excited states.
Table~\ref{tab:ex_dw} shows the summary of calculations.
Basically, not only the ground-state but also low-lying excited-states wave functions and energies are successfully calculated,
even for the degenerate states.
It is not apparent which calculation gives, left-right bases ($ \psi_{\urm{L}} $ and $ \psi_{\urm{R}} $),
parity bases ($ \psi_{+} $ and $ \psi_{-} $),
or even general linear combinations.
The DNN calculations for both $ \alpha = 2.0 $ and $ 3.0 $ obtained wave functions with the left-right bases,
while it may depend on the initial condition.
Note that the exact diagonalization for $ \alpha = 2.0 $ obtained wave functions with the parity bases,
while wave functions with the left-right bases are plotted by using linear combinations in Fig.~\ref{fig:ex_dw_wf_2} to make a comparison with the DNN result easily.
\begin{table*}[tb]
  \centering
  \caption{
    Calculation summary of a one-body problem under the double-well potential.}
  \label{tab:ex_dw}
  \begin{ruledtabular}
    \begin{tabular}{dldddd}
      \multicolumn{1}{c}{$ \alpha $} & \multicolumn{1}{c}{$ j $} & \multicolumn{2}{c}{Energy} & \multicolumn{1}{c}{Epochs} & \multicolumn{1}{c}{Time per Epoch ($ \mathrm{\mu s} $)} \\
      \cline{3-4}
                                     & & \multicolumn{1}{c}{Exact diagonalization} & \multicolumn{1}{c}{Deep neural network} & & \\

      \hline
      1.0 & 0 & +0.869573 & +0.869706 & 29108 & 513.203 \\
      1.0 & 1 & +1.661393 & +1.661685 & 35596 & 523.556 \\
      1.0 & 2 & +3.543667 & +3.544327 & 55037 & 524.351 \\
      1.0 & 3 & +5.665058 & +5.666010 & 20083 & 536.858 \\
      \hline      
      1.25 & 0 & +1.417858 & +1.417886 & 30284 & 512.083 \\
      1.25 & 1 & +1.725904 & +1.726013 & 36992 & 523.092 \\
      1.25 & 2 & +3.717933 & +3.717949 & 30943 & 528.675 \\
      1.25 & 3 & +5.424725 & +5.424850 & 24474 & 534.907 \\
      \hline
      2.0 & 0 & +2.762317 & +2.762333 & 23663 & 519.261 \\
      2.0 & 1 & +2.762333 & +2.762343 & 24764 & 529.129 \\
      2.0 & 2 & +7.988520 & +7.989654 & 19491 & 532.193 \\
      2.0 & 3 & +7.990618 & +7.989601 & 18662 & 534.543 \\
      \hline
      3.0 & 0 & +4.214229 & +4.214253 & 20526 & 515.042 \\
      3.0 & 1 & +4.214229 & +4.214284 & 20290 & 521.291 \\
      3.0 & 2 & +12.526202 & +12.526214 & 51131 & 519.062 \\
      3.0 & 3 & +12.526202 & +12.526278 & 12488 & 534.752 \\
    \end{tabular}
  \end{ruledtabular}
\end{table*}
\subsection{Two-body systems}
\par
Two-body one-dimensional harmonic oscillators are taken as the last examples.
Here, $ x_{\urm{max}} = y_{\urm{max}} = 5 $ and $ M_x = M_y = 256 $ are used for the spatial mesh
and
two layers each of which contains $ 32 $ units are used for the DNN.
Here, the interparticle interaction is not considered.
The exact wave functions for several low-lying excited states can be written as
linear combinations of Eqs.~\eqref{eq:ho_ex}:
\begin{subequations}
  \begin{align}
    \Psi_0 \left( x, y \right)
    & =
      \psi_0 \left( x \right)
      \psi_0 \left( y \right), \\
    \Psi_1 \left( x, y \right)
    & =
      \frac{1}{\sqrt{2}}
      \left[
      \psi_0 \left( x \right)
      \psi_1 \left( y \right)
      +
      \psi_1 \left( x \right)
      \psi_0 \left( y \right)
      \right], \\
    \Psi_2 \left( x, y \right)
    & =
      \frac{1}{\sqrt{2}}
      \left[
      \psi_0 \left( x \right)
      \psi_2 \left( y \right)
      +
      \psi_2 \left( x \right)
      \psi_0 \left( y \right)
      \right], \\
    \Psi_3 \left( x, y \right)
    & = 
      \psi_1 \left( x \right)
      \psi_1 \left( y \right),
  \end{align}
\end{subequations}
where the energy eigenvalue of $ \Psi_0 $, $ \Psi_1 $, $ \Psi_2 $, and $ \Psi_3 $ are equal to, respectively,
$ 1 $, $ 2 $, $ 3 $ and $ 3 $ 
for bosonic systems
and 
\begin{subequations}
  \begin{align}
    \Psi_0 \left( x, y \right)
    & =
      \frac{1}{\sqrt{2}}
      \begin{vmatrix}
        \psi_0 \left( x \right) & \psi_1 \left( x \right) \\
        \psi_0 \left( y \right) & \psi_1 \left( y \right) 
      \end{vmatrix}, \\
    \Psi_1 \left( x, y \right)
    & =
      \frac{1}{\sqrt{2}}
      \begin{vmatrix}
        \psi_0 \left( x \right) & \psi_2 \left( x \right) \\
        \psi_0 \left( y \right) & \psi_2 \left( y \right) 
      \end{vmatrix}, \\
    \Psi_2 \left( x, y \right)
    & =
      \frac{1}{\sqrt{2}}
      \begin{vmatrix}
        \psi_0 \left( x \right) & \psi_3 \left( x \right) \\
        \psi_0 \left( y \right) & \psi_3 \left( y \right) 
      \end{vmatrix}, \\
    \Psi_3 \left( x, y \right)
    & =
      \frac{1}{\sqrt{2}}
      \begin{vmatrix}
        \psi_1 \left( x \right) & \psi_2 \left( x \right) \\
        \psi_1 \left( y \right) & \psi_2 \left( y \right) 
      \end{vmatrix},
  \end{align}
\end{subequations}
where the energy eigenvalue of $ \Psi_0 $, $ \Psi_1 $, $ \Psi_2 $, and $ \Psi_3 $ are equal to, respectively,
$ 2 $, $ 3 $, $ 4 $ and $ 4 $ 
for fermionic systems.
Note that the second excited states, $ \Psi_2 $ and $ \Psi_3 $, are twofold degenerate
in both the bosonic and fermionic systems.
\par
\begin{figure*}[tb]
  \centering
  \includegraphics[width=1.0\linewidth]{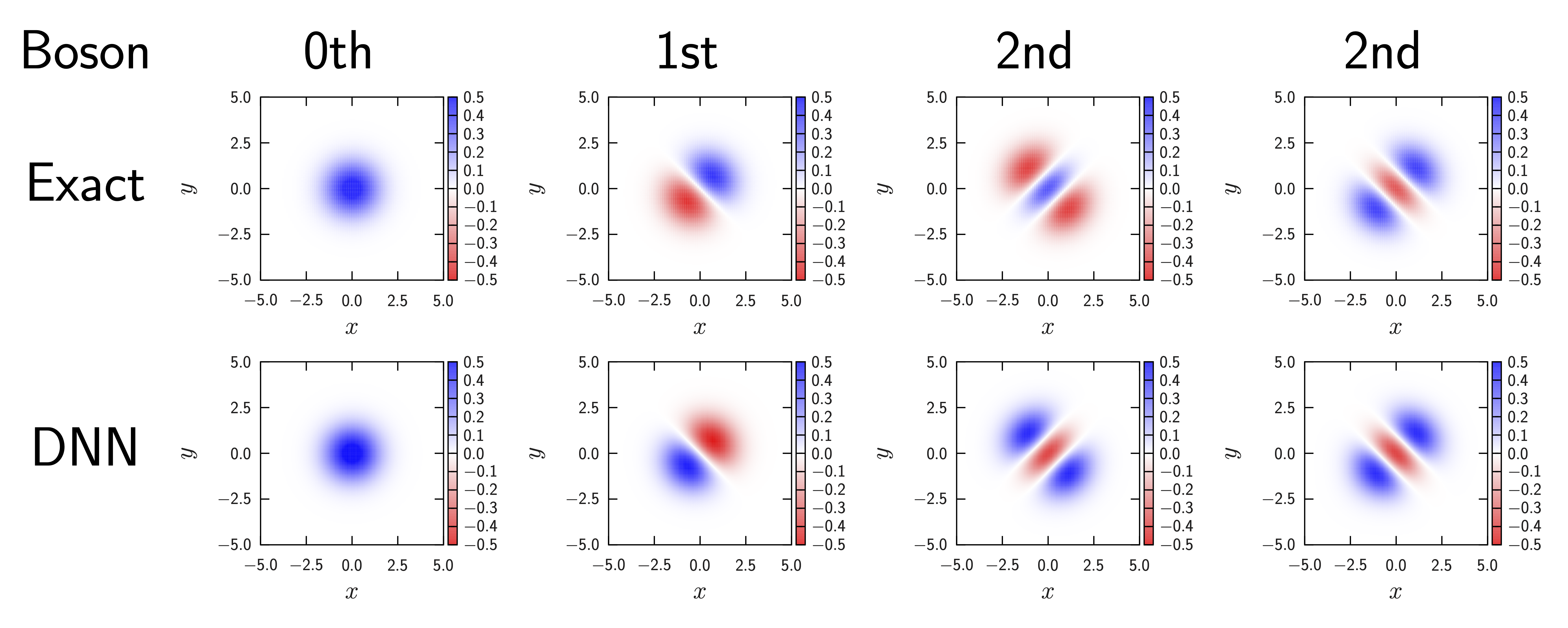}
  \caption{
    Two-body wave function for the ground and low-lying excited states under the harmonic oscillator potential without the interaction for bosonic systems.
    The exact wave function is shown in the top row.}
  \label{fig:2body_ho_wf_ex_boson}
\end{figure*}
\begin{figure*}[tb]
  \centering
  \includegraphics[width=1.0\linewidth]{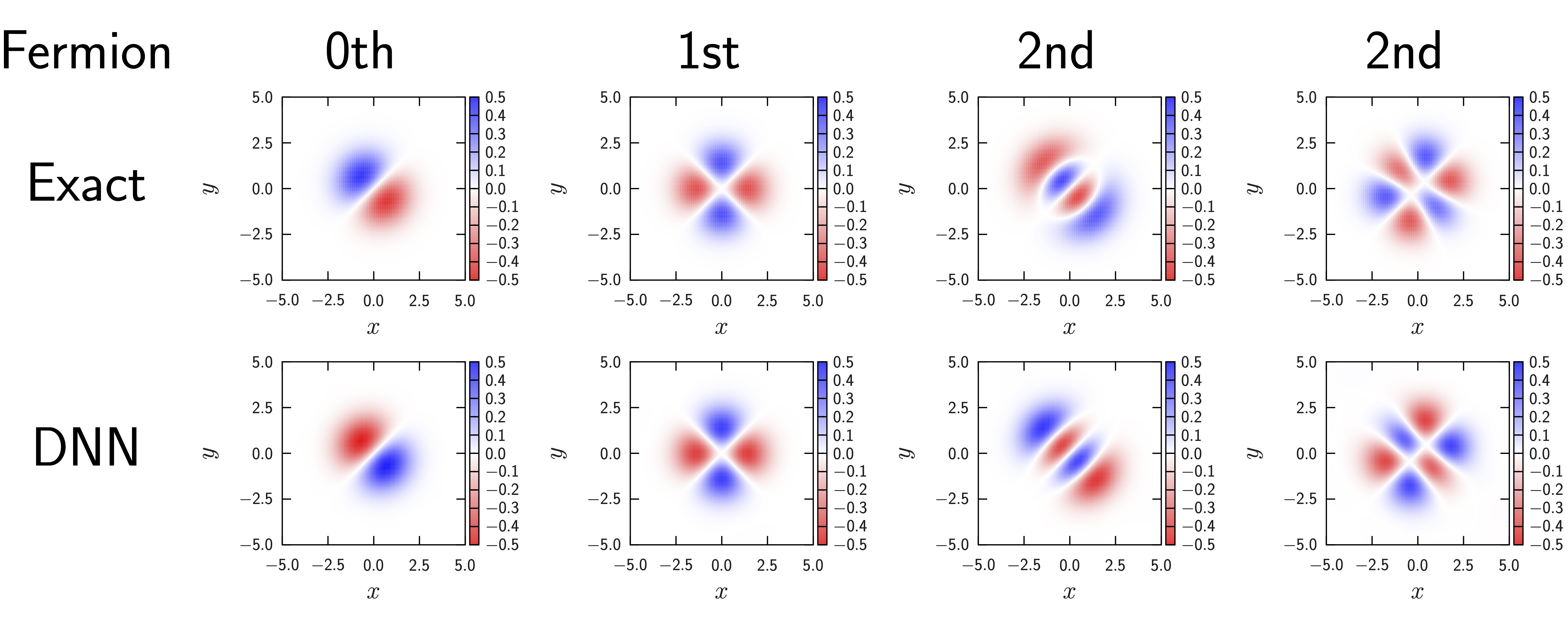}
  \caption{
    Same as Fig.~\ref{fig:2body_ho_wf_ex_boson} but for fermionic systems.}
  \label{fig:2body_ho_wf_ex_fermion}
\end{figure*}

Figures~\ref{fig:2body_ho_wf_ex_boson} and \ref{fig:2body_ho_wf_ex_fermion},
respectively,
show the wave functions of the ground state and
first and second excited states.
Table~\ref{tab:ex_2body_ho} shows the summary of calculations.
Note that
$ \left[ \Psi_2 \left( x, y \right) \pm \Psi_3 \left( x, y \right) \right] / \sqrt{2} $
are plotted for the second excited states for exact solutions.
Not only the ground-state but also low-lying excited-states wave functions and energies are successfully calculated
even for two-body systems.
In addition, as one-body problems,
the numerical cost for a low-lying excited state is almost the same as that for the ground-state.
Thus, this method to calculate low-lying excited states can work even for multibody systems with a reasonable numerical cost.
\begingroup
\squeezetable
\begin{table}[tb]
  \centering
  \caption{
    Calculation summary of excited states for a two-body problem under the harmonic oscillator potential.
    Calculation is performed with $ \omega = 1.0 $.}
  \label{tab:ex_2body_ho}
  \begin{ruledtabular}
    \begin{tabular}{llddd}
      \multicolumn{1}{c}{Particles} & \multicolumn{1}{c}{State} & \multicolumn{1}{c}{Energy} & \multicolumn{1}{c}{Epochs} & \multicolumn{1}{c}{Time per Epoch ($ \mathrm{ms} $)} \\
      \hline
      Boson & 0th & +0.999935 & 26197 & 23.859 \\
      Boson & 1st & +1.999778 & 26224 & 24.228 \\
      Boson & 2nd (1) & +2.999801 & 19204 & 24.502 \\
      Boson & 2nd (2) & +3.000229 & 15239 & 25.450 \\
      \hline
      Fermion & 0th & +1.999855 & 38500 & 23.844 \\
      Fermion & 1st & +2.999771 & 29552 & 24.205 \\
      Fermion & 2nd (1) & +3.999251 & 28197 & 24.674 \\
      Fermion & 2nd (2) & +4.004341 & 11871 & 25.772 \\
    \end{tabular}
  \end{ruledtabular}
\end{table}
\endgroup
%
% Conclusion
%
\section{Summary}
\label{sec:summary}
\par
In this paper, we proposed a method to calculate
the wave functions and energies of 
not only the ground state but also low-lying excited states
of quantum multibody systems
using the deep neural network and the unsupervised machine learning technique.
To calculate systems of many-particle systems of identical particles,
a simple method of 
symmetrization for bosonic systems
and
antisymmetrization for fermionic systems
were also proposed.
\par
The obtained wave functions and energies are consistent with the exact solution.
We found that the neural network is not necessarily large for one-body systems,
which also enables us to analyze the internal structure of the deep neural network used.
For instance, just only one hidden layer with four units is enough to describe the ground-state wave function of the harmonic oscillator.
This can be understood by using the piecewise approximation with linear functions.
We confirmed that our simple (anti)symmetrization method works for multibody systems.
The numerical cost per epoch for fermionic systems is almost the same as that for bosonic systems.
The numerical cost is almost proportional to the number of spatial meshes since the sparse matrix representation is used.
In addition, the numerical cost for a low-lying excited state is almost the same as that for the ground state.
\par
The deep neural network has been applied to solve many-fermion systems
where the ground-state wave function is assumed to be a Jastrow wave function~\cite{
  Pfau2020Phys.Rev.Research2_033429,
  Hermann2020Nat.Chem.12_891,
  Entwistle2023Nat.Commun.14_274}.
The method proposed in this paper can be an alternative method to solve many-fermionic systems
since the ansatz for the ground-state wave function is lenient,
and the symmetrization and antisymmetrization are treated on an equal footing.
\par
Since the numerical cost is not so large and our (anti)symmetrization is quite simple,
this method can be an alternative method to calculate wave functions and energies of the ground and low-lying excited states,
for instance,
for the electronic structure of molecules and solids,
for the nuclear structure of atomic nuclei
including a tetra neutron~\cite{
  Kisamori2016Phys.Rev.Lett.116_052501,
  Faestermann2022Phys.Lett.B824_136799,
  Duer2022Nature606_678},
and for cold atoms~\cite{
  Wenz2013Science342_457}.
\par
At this moment, we only considered one-dimensional systems,
while most problems interested are three-dimensional systems.
In addition, spin components, or even isospin components for nuclear systems, are often important.
The restricted Boltzmann machine has been applied to obtain the ground- and low-lying excited-states wave functions~\cite{
  Nomura2017Phys.Rev.B96_205152,
  Choo2018Phys.Rev.Lett.121_167204,
  Nomura2020J.Phys.Soc.Japan89_054706}.
Since the input is discrete variables in the spin systems, 
the Boltzmann machine is suitable.
Such pioneering works may help to consider the spin (or isospin) components in this work.
Such extensions are possible within our framework, and remain for future work.
\par
As far as we know,
all the calculations using the deep neural network for wave functions
are static,
while describing many phenomena including
the interaction between matter and laser~\cite{
  Sato2015Phys.Rev.B92_205413,
  Tanaka2022J.Phys.Cond.Matt.34_165901},
ion-cluster collision~\cite{
  Yabana1998Phys.Rev.A57_R3165}
heavy-ion collision~\cite{
  Sekizawa2013Phys.Rev.C88_014614}
nuclear fission~\cite{
  Back2014Rev.Mod.Phys.86_317,
  Zhao2019Phys.Rev.C99_054613},
and fusion~\cite{
  Sekizawa2019Phys.Rev.C99_051602}.
To describe such phenomena,
time evolution from a state obtained by the deep neural network is also interesting,
while it is left for a future study.
\par
Finally, let us make a comment on the interpretation of the wave functions obtained in our work.
As we have shown, thanks to the simplicity of the deep neural network,
we could interpret the structure of the network easily.
We found that replacing the softmax function
with the ReLU activation provides a piecewise linear function which approximates the ground-state wave function.
Since any wave function including those for excited states, which is naturally continuous, 
can be approximated by a piecewise-linear function, we intuitively conclude that
the neural network representation can work for any physical quantum mechanical system in any dimensions.
The physical meaning of the piecewise-linear functions is as follows.
First of all, linear functions are solutions 
of the free Schr\"{o}dinger equation with no potential term.
So it is a good idea to start with linear functions in
physical systems.
Then the inclusion of the potential term in the Hamiltonian causes the curvature of the
wave function.
The curvature is determined by the interplay between the Laplacian and the potential term in the Hamiltonian.
So, the kink structure of the wave function is dictated by the Hamiltonian.
The kinks correspond to the ReLU activations, thus in effect, the nonlinearity in the Hamiltonian corresponds
to the neural network structure.
This reminds us of the work~\cite{
  carleo2018constructing}
in which the deep layers 
of the deep Boltzmann machine representing the ground-state wave functions of spin systems 
were interpreted as a Euclidean Hamiltonian evolution,
or the work~\cite{
  hashimoto2018deep,
  hashimoto2018deep2,
  hashimoto2019ads} 
in which the deep layers of the sparse neural network used for the AdS/CFT correspondence
were interpreted as a bulk curved geometry.
Further interplay between the sparsity of the 
interpretable neural networks and Hamiltonians of physical systems is to be discovered.
%
% Acknowledgement
%
\begin{acknowledgments}
  The authors acknowledge the fruitful discussion with
  Haozhao Liang,
  Masaaki Kimura,
  and
  Hiroyuki Tajima.
  T.~N.~also thanks
  Chuanxin Wang for the careful reading of the manuscript
  and
  Yuki Nagai for pointing out the typo of Eq.~\eqref{eq:interaction}.
  T.~N.~acknolwedges
  the RIKEN Special Postdoctoral Researcher Program,
  the Science and Technology Hub Collaborative Research Program from RIKEN Cluster for Science, Technology and Innovation Hub (RCSTI),
  the JSPS Grant-in-Aid for Research Activity Start-up under Grant No.~JP22K20372,
  the JSPS Grant-in-Aid for Transformative Research Areas (A) under Grant No.~JP23H04526,
  the JSPS Grant-in-Aid for Scientific Research (B) under Grant No.~JP23H01845,
  and
  the JSPS Grant-in-Aid for Scientific Research (C) under Grant No.~JP23K03426.
  H.~N.~acknowledges
  the JSPS Grant-in-Aid for Scientific Research (C) under Grant No.~JP19K03488
  and
  the JSPS Grant-in-Aid for Scientific Research (B) under Grant No.~JP23H01072.
  The work of K.~H.~was supported in part by JSPS KAKENHI Grants No.~JP22H01217, No.~JP22H05111 and No.~JP22H05115.
\end{acknowledgments}
%
%%%%%%%%%%%%%%%%%%%%%%%%%%%%%%%%%%%%%%%%%%%%%%%%%% 
% 
\bibliography{024_dnnqm_antisym}
% 
%%%%%%%%%%%%%%%%%%%%%%%%%%%%%%%%%%%%%%%%%%%%%%%%%%
\end{document}